\documentclass[fleqn,10pt]{wlscirep}
\usepackage[utf8]{inputenc}
\usepackage[T1]{fontenc}
\usepackage{bm}
\usepackage{comment}

\usepackage{amsmath}
\usepackage{color}
\usepackage{lineno}
\title{Low power continuous-wave all-optical magnetic switching in ferromagnetic nanoarrays} 

\author[1,3,*]{Kilian D. Stenning}
\author[1,3]{Xiaofei Xiao}
\author[1]{Holly H. Holder}
\author[1]{Jack C. Gartside}
\author[1,2]{Alex Vanstone}
\author[1,2]{Oscar W. Kennedy}
\author[1]{Rupert F. Oulton}
\author[1]{Will R. Branford}
\affil[1]{Blackett Laboratory, Imperial College London, London SW7 2AZ, United Kingdom}
\affil[2]{London Centre for Nanotechnology, University College London, London WC1H 0AH, United Kingdom}
\affil[3]{These authors contributed equally}
\affil[*]{Corresponding author e-mail: k.stenning18@imperial.ac.uk}

\begin{abstract}

All-optical magnetic switching promises ultrafast, high-resolution magnetisation control with the technological attraction of requiring no magnetic field. Existing all-optical switching schemes are driven by ultrafast transient effects, typically requiring power-hungry femtosecond-pulsed lasers and complex magnetic materials. Here, we demonstrate deterministic, all-optical magnetic switching in simple ferromagnetic nanomagnets (Ni$_{81}$Fe$_{19}$, Ni$_{50}$Fe$_{50}$) with sub-diffraction limit dimensions using a focused low-power, linearly-polarised continuous-wave laser. Isolated nanomagnets are switched across a range of dimensions, laser wavelengths and powers. All square-geometry artificial spin ice vertex configurations are written, including ground-state and energetically-unfavourable `monopole-like' states at powers as low as 2.74 mW. Usually, magnetic switching with linearly polarised light is symmetry-forbidden; however, here the laser spot has a similar size to the nanomagnets, producing an absorption distribution dependent on the relative nanoisland-spot displacement. We attribute the observed deterministic switching to the transient dynamics of this asymmetric absorption. No switching is observed in Co samples, suggesting the multi-species nature of NiFe alloys plays a role in reversal. The results presented here usher in cheap, low-power optically-controlled devices with impact across data storage, neuromorphic computation and reconfigurable magnonics.

\end{abstract}

\begin{document}

\flushbottom
\maketitle
\thispagestyle{empty}

\section*{Introduction}
Efficient optical control of magnetic materials is a long-standing goal for data storage and computational technologies. Continuous-wave exposure of magnetic materials can have a number of effects, from linearly-polarised light induced modifications of the magnetic anisotropy, susceptibility and coercivity to magnetisation control via angular momentum transfer from circularly polarised light\cite{kovalenko1986photoinduced}. The latter is promising for data storage, yet in ferromagnets, domains can only be grown (due to magnetic circular dichroism effects) and not switched. 75$\%$ of global data is stored magnetically and the predominant recording technology uses power-consuming magnetic fields with a plasmonically-focused laser beam for Heat Assisted Magnetic Recording (HAMR)\cite{pancaldi2019selective,kryder2008heat}. Removing magnetic field requirements for all-optical magnetic switching (AOMS) \cite{stanciu2007all,kirilyuk2010ultrafast,kimel2019writing} represents a next-generation class of local magnetisation control, with intriguing switching effects beyond current thermodynamic descriptions and wide-ranging technological implications.

AOMS has been demonstrated in several schemes via ultrafast excitation with varying underlying switching mechanisms. Helicity-dependent switching (HDS) with single or multiple femtosecond laser pulses has been observed in ferrimagnetic, ferromagnetic and granular magnetic media\cite{stanciu2007all,el2016two,  medapalli2017multiscale,ellis2016all,lambert2014all,kichin2019multiple}. In these systems, light serves to both demagnetise the material and realign the magnetisation via angular momentum transfer (arising from the inverse Faraday effect or magnetic circular dichroism). The effective field breaks system symmetry, driving magnetic reversal. 

Helicity-independent switching (HIS) has been demonstrated with a single femtosecond pulse, yet is largely limited to rare-earth ferrimagnets, typically GdFeCo\cite{ostler2012ultrafast,stanciu2007subpicosecond,radu2011transient,el2016two,el2017materials,le2012demonstration}. In these systems, the electron and spin temperatures and the different band structures (with different demagnetisation rates) between sublattices play a crucial role. As the system is excited, there is a marked increase in the electron and spin temperatures\cite{ostler2012ultrafast}. This leads to the FeCo sublattice rapidly demagnetising within $<$ 1 ps, as observed in a number of ferromagnetic materials\cite{beaurepaire1996ultrafast}. Due to the different magnetic moment, Gd demagnetises at a slower rate, and partial angular momentum transfer to the remagnetising FeCo results in a transient ferromagnetic state\cite{radu2011transient}. As the magnetisation and temperatures equilibrate, the magnetisation reverses. The same effect is observed with hot electron injection\cite{xu2017ultrafast}, confirming the thermal nature of the theory. HIS has also been observed in rare-earth free ferrimagnets with near-identical sublattice demagnetisation rates\cite{banerjee2020single} where switching is believed to be driven by exchange scattering and angular momentum transfer.

HIS in ferromagnets is typically observed when coupled to GdFeCo layers \cite{gorchon2017single,iihama2018single,igarashi2020engineering} where spin-polarised currents generated from the HIS of GdFeCo transfer angular momentum to the demagnetised ferromagnetic layer causing reversal\cite{igarashi2020engineering}. HIS has been demonstrated in Pt/Co/Pt multilayers through a thermal process when focusing the laser spot to single-domain sizes\cite{vomir2017single}, implying the boundary of the excitation plays an important role. These demonstrations have shown the promise of AOMS, but the requirement of huge femtosecond-pulsed MW lasers and exotic magnetic materials renders them unsuitable for device integration and up-scaling for application. Additionally, the majority of studies concern continuous thin films or well-spaced single nanostructures\cite{le2012demonstration}, restricting write-density.

Artificial spin systems comprising networks of strongly-interacting nanomagnets serve as promising hosts for future information-processing technologies including nanomagnetic logic\cite{camsari2017stochastic,caravelli2020logical}, neuromorphic computation\cite{tanaka2019recent,markovic2020physics,gartside2021reservoir,jensen2018computation,jensen2020reservoir,hon2021numerical} and reconfigurable magnonics\cite{lendinez2019magnetization,chumak2017magnonic,barman2020magnetization,kaffash2021nanomagnonics,gartside2021reconfigurable,stenning2020magnonic,gartside2020current,dion2022observation}. Information can be stored in the magnetisation of a single nanomagnet or the magnetic configuration of the entire network (microstate), where collective microstate-dependent dynamics\cite{lendinez2019magnetization, vanstone2021spectral, gartside2021reconfigurable} may be harnessed to process information\cite{jensen2018computation,jensen2020reservoir,hon2021numerical,gartside2021reservoir}. Local nanomagnet switching has been achieved through diffraction-limited heat-assisted reversal relying on global fields in conjunction with laser illumination\cite{pancaldi2019selective,gypens2021thermoplasmonic} and by cumbersome field-assisted\cite{wang2016rewritable} and field-free scanning-probe\cite{gartside2016novel,gartside2018realization,stenning2020magnonic} techniques. In the latter, the magnetic field from a magnetic tip breaks the nanomagnet symmetry, causing transient domain-wall formation and asymmetric propagation leading to switching\cite{gartside2018realization}. Realising rapid low-power single-element magnetic switching is pivotal for the development of functional nanomagnetic computation and storage devices.

Here, we demonstrate deterministic all-optical magnetic switching in both isolated nanomagnets and dense square artificial spin ice (ASI)\cite{wang2006artificial, skjaervo2020advances,lendinez2019magnetization} arrays using simple NiFe alloys and nanofocused low-power ($\sim$2.74 mW), linearly-polarised continuous-wave (CW) lasers with no external magnetic fields applied. Under certain conditions, the optical excitation leads to an asymmetric absorption across the nanomagnet, breaking the system symmetry and providing a driving force for switching. Despite the long-exposure times ($\sim$ms) a single, deterministic `toggle' switch, or magnetic inversion, occurs. This is distinct from helicity-dependent switching, which is magnetic-field-like and writes a specific magnetisation direction, implying our observed switching is a result of transient dynamics as the system approaches thermodynamic equilibrium. The substrates (here Si/SiO\textsubscript{2} or Au/SiO\textsubscript{2}) on which the nanomagnets are situated have an anti-reflection function, which enhances optical absorption within the nanomagnets up to 65$\%$ of the total incident light across a broad frequency range. Sub-diffraction `writing' and `erasing' of information is achieved in single nanomagnets. All square-ASI vertex configurations are deterministically written, from low-energy ground state to high-energy monopole-like states, irrespective of their dipolar energies - ruling out thermal relaxation. Switching is not observed in Co nanostructures, suggesting multi-species interactions play a role in the reversal mechanism. Our results open up intriguing questions into the manipulation of nanostructured ferromagnets with linearly-polarised light and usher in a new paradigm of low-power magnetisation control with vast implications across data-storage, computation and magnonics.

\subsection*{Deterministic reversal of isolated nanomagnets}

\begin{figure}[tb]
    \centering
    \includegraphics[width=\textwidth]{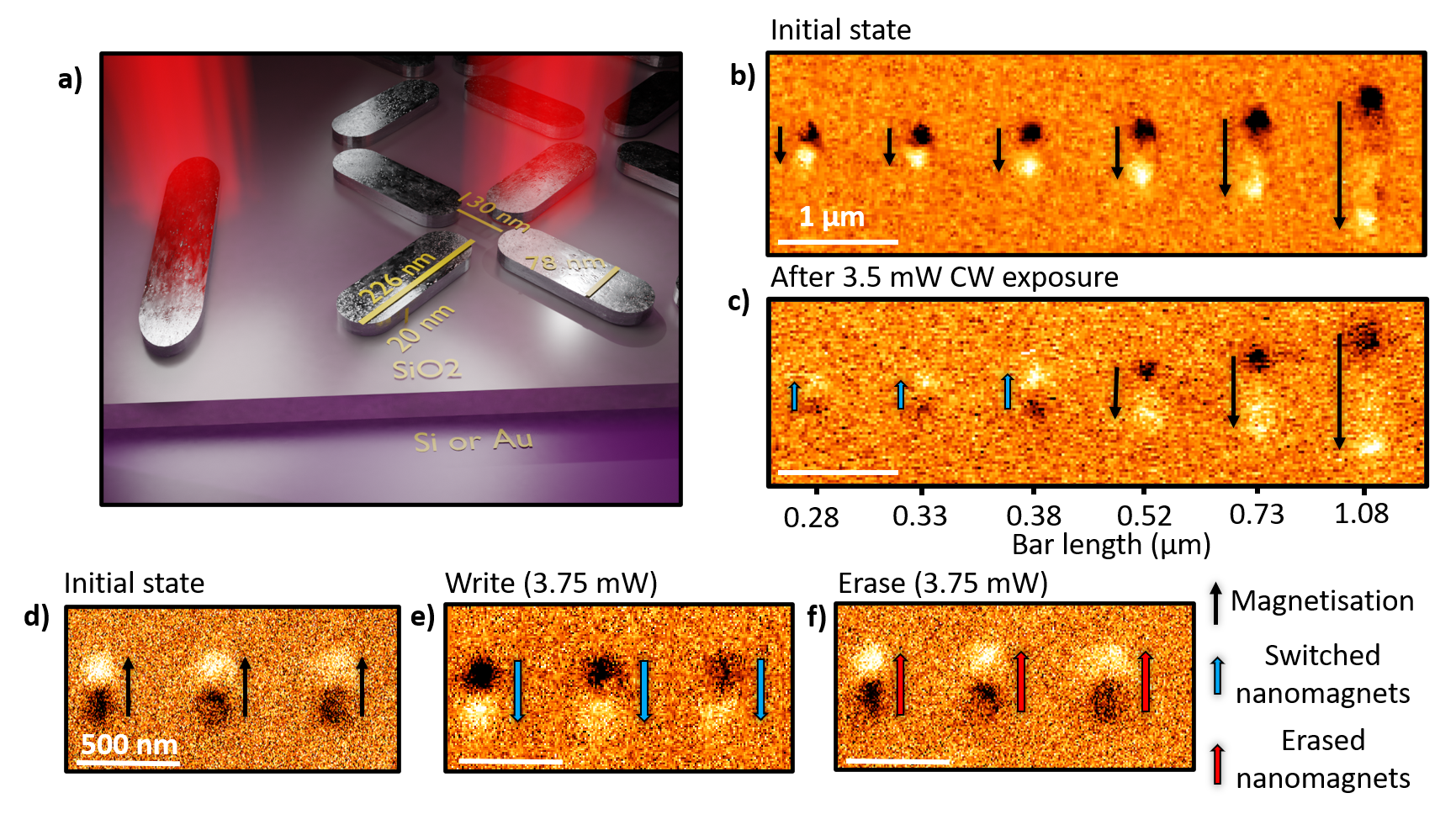}
    \caption{a) Schematic of linearly-polarised CW laser exposure of isolated nanomagnets (left) and densely-packed Py (NiFe) nanomagnets in a square ASI (right) patterned on Si or Au substrates with a SiO$_2$ coating.  MFM images of isolated nanomagnets patterned on an Au (250 nm) / SiO$_2$ (290 nm) substrate b) in the saturated state and c) after linearly-polarised CW laser exposure at 3.5 mW power.  Bars with length (L) $\leq$ 380 nm switch.  MFM images of d) three nanomagnets with L = 280 nm in an initial saturated state. e) Subsequent switching after linearly-polarised CW exposure with 3.75 mW. f) Switching after a second 3.75 mW exposure demonstrating the ability to rewrite / erase previously written states. In each case laser polarisation is parallel to long-axis of nanomagnets and the laser scans perpendicular to the bar long axis.}
    \label{FTIR}
\end{figure}

Figure \ref{FTIR} a) shows a schematic of nanomagnets exposed with a focused linearly-polarised CW laser. We first consider reversing isolated Ni$_{81}$Fe$_{19}$ (permalloy, Py) nanomagnets separated by 1 \textmu m such that dipolar interactions are negligible. The nanomagnets are patterned on top of an Au (250 nm) / SiO$_2$ (290 nm) substrate which displays enhanced optical absorption (see supplementary note 1). Figure \ref{FTIR} b) shows a `before' magnetic force microscopy (MFM) image of field-saturated nanomagnets with lengths $L$ = 0.28-1.08 \textmu m. Each bar exhibits a positive (light) and negative (dark) magnetic charge indicating magnetisation direction, shown by adjacent arrows. Figure \ref{FTIR} c) shows the same bars after exposure to a $\lambda$ = 633 nm linearly-polarised (parallel to nanomagnet long-axis) CW laser with 580 nm diameter spot and 3.5 mW power, swept across the nanomagnets perpendicular to their long axis. Magnetic switching is observed in the three left-most bars (L $\leq$ 0.28-0.38 \textmu m), indicated by light-blue arrows. Figures \ref{FTIR} d-f) demonstrate write/erase functionality in three L = 0.28 \textmu m bars. Bars are initially globally-saturated (Figure  \ref{FTIR} d)), then exposed to a $\lambda$ = 633 nm, 3.75 mW power linearly-polarised CW laser such that all bars switch (`write', Figure  \ref{FTIR} e)). All bars then switch back following a second exposure (`erase', Figure  \ref{FTIR} f)). The observation of toggle-switching functionality implies the underlying mechanism is not magnetic-field-like, as repeating the same write operation inverts the magnetisation each time. Note that powers stated are measured before focussing and $\leq$ 90$\%$ of the power reaches the sample (see methods for details).

\subsection*{Deterministic reversal in ASI networks}

\begin{figure}[tb!]
    \centering
    \includegraphics[width=0.8\textwidth]{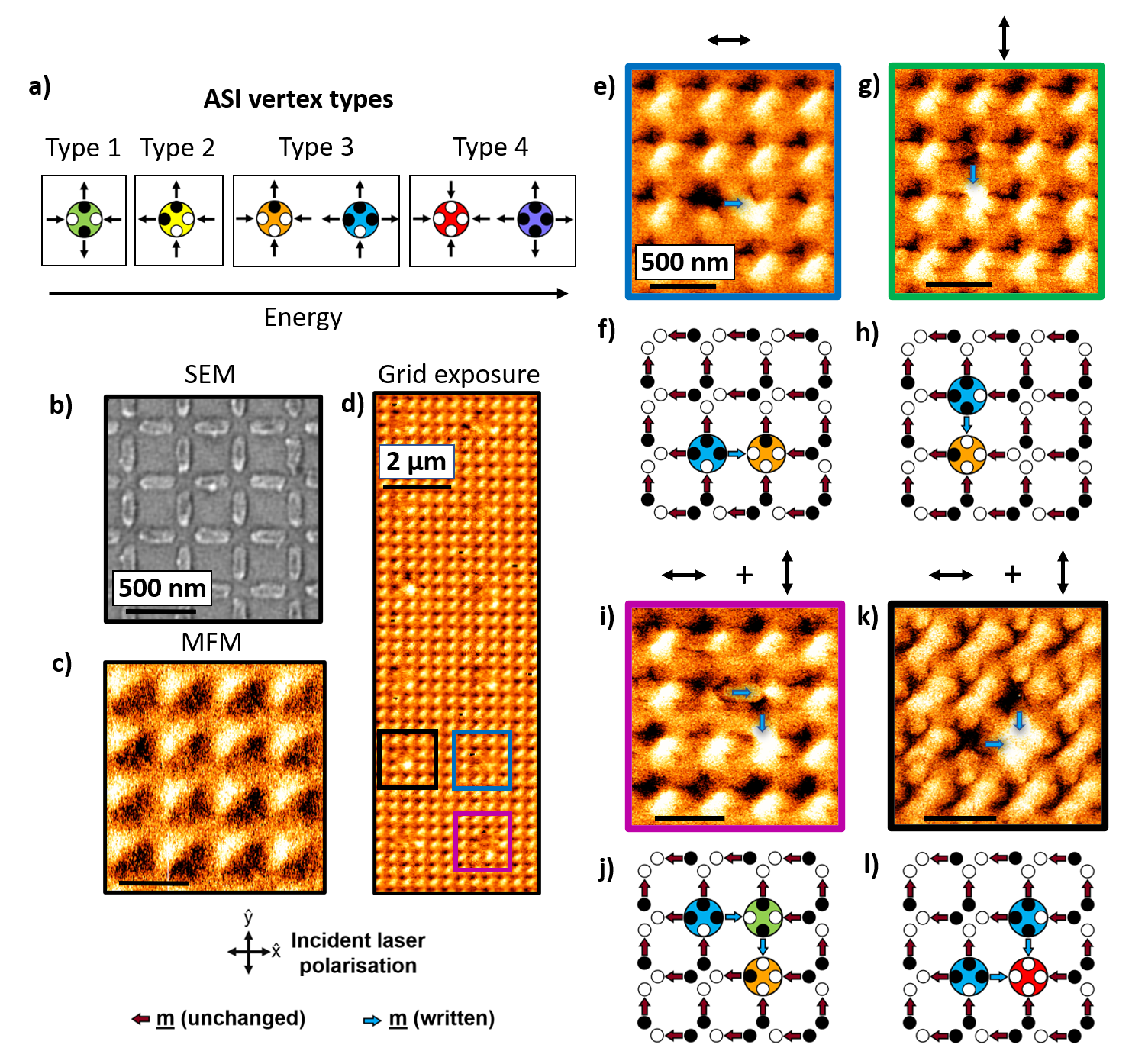}
    \caption{Single nanomagnet switching in a dense array. a) Vertex types in square ASI. b) SEM image of a square ASI array. c) MFM images of an equivalent array after global-field initialisation. d) MFM image showing a grid of reversals after $\lambda$ = 663 nm laser exposure with a 580 nm focal spot at 5 mW for $\sim$ 0.25 s. Zoomed-in MFM images and schematics of colour-coded regions of panel d) showing nanomagnet reversal after laser exposure with e,f) $\hat{x}$ polarisation, g,h) $\hat{y}$ polarisation, i-l) $\hat{x}$ then $\hat{y}$ polarisation. Panel g) is from a different region of the array. The single polarisation exposure in e-h) result in a pair of Type 3 vertices. When exposing both polarisations, both i,j) Type 1 (GS) and k,l) high-energy T4 (MP) vertices can be prepared depending on which nanomagnets are exposed. In e-l) the scale bar corresponds to 500 nm. Only written vertices are coloured.}
    \label{dots}
\end{figure}
We now further explore the exposure of strongly-interacting, sub-diffraction limited nanomagnets arranged in a dense square ASI, patterned on a Si / SiO\textsubscript{2} (300 nm) substrate. This substrate was selected to demonstrate the efficacy of the switching technique on commercially available, sub-optimal substrates (no additional absorption-enhancing Au layer, supplementary note 1). Strong interactions between neighbouring nanomagnets lead to four vertex types with differing energies\cite{wang2006artificial} (Figure  \ref{dots}a)), providing a challenging proving ground for local magnetic control as strong dipolar interactions will oppose writing of high-energy states. Figure \ref{dots} a) shows all possible vertex configurations in square ASI in order of increasing energy, labelled as Types 1-4 (vertex energies detailed in supplementary note 4). Figure \ref{dots} b) shows an SEM image of the ASI array. Nanomagnet dimensions are $L = 226$ nm, $W = 78$ nm and $P = 356$ nm. 

First, we consider exposing ASI to discrete spot illuminations of a $\lambda$ = 633 nm, 5 mW CW laser for $\sim$ 0.25 s. The laser position is static throughout illumination. Laser focal spot size is 580 nm. Prior to exposure, the ASI is initialised via global-field saturation (-$x$,+$y$) to define an array of type 2 vertices (Figure \ref{dots} c)). Figure \ref{dots} d) shows the array microstate after grid like spot exposures with a variety of laser polarisations. Laser polarisation governs the resultant written vertex state, with colour-coded regions of Figure \ref{dots} d) shown zoomed-in in Figure \ref{dots} e-l). Higher powers and longer illuminations lead to sample damage (not shown). The zoomed regions show successful writing of ASI vertex types 1 (i,j), 3 (e,f) and 4 (k,l) (g,h are from a different region of the array). Note that the MFM images are supplemented with magnetic charge schematics and the corresponding laser polarisations required to prepare each state, indicated above each MFM image. When illuminating the array at a single spot (Figure \ref{dots} e-h)), a single nanomagnet with long-axis parallel to the laser polarisation switches. This selectivity arises from the polarisation-dependent absorption (see supplementary note 1). This forms a high-energy state comprising a pair of type 3 vertices. Two subsequent $\hat{x}$ then $\hat{y}$-polarised exposures results in two nanomagnets switching (one from each $x,y$ subset). Here both low energy Type 1 (Figure \ref{dots} i,j)) and high-energy Type 4 (Figure \ref{dots} k,l)) vertices may be written. The position of the incident laser with respect to the surrounding microstate determines whether a Type 1 or Type 4 is written. The ability to write type 4 vertices strongly suggests that switching occurs in a deterministic manner rather than via nanomagnet thermalisation and relaxation, which would strongly-favour the system's ground state and that dipolar interactions does not affect switching.

This Gaussian distribution of power across the beam gives a FWHM of $\sim$ 340 nm. Combining this with selectivity enabled by the polarisation-dependent absorption predominantly results in a single nanomagnet switching during each exposure. We observe a $\sim$ 20 \% chance of switching two nanomagnets with the same polarisation which may be reduced by using larger nanomagnets or a smaller design wavelength. The observed writing of low and high-energy states with single-nanomagnet precision shows a deterministic reversal mechanism and illustrates the applicability of this method to next-generation storage and memcomputing.

\begin{figure}[t!]
    \centering
    \includegraphics[width=\textwidth]{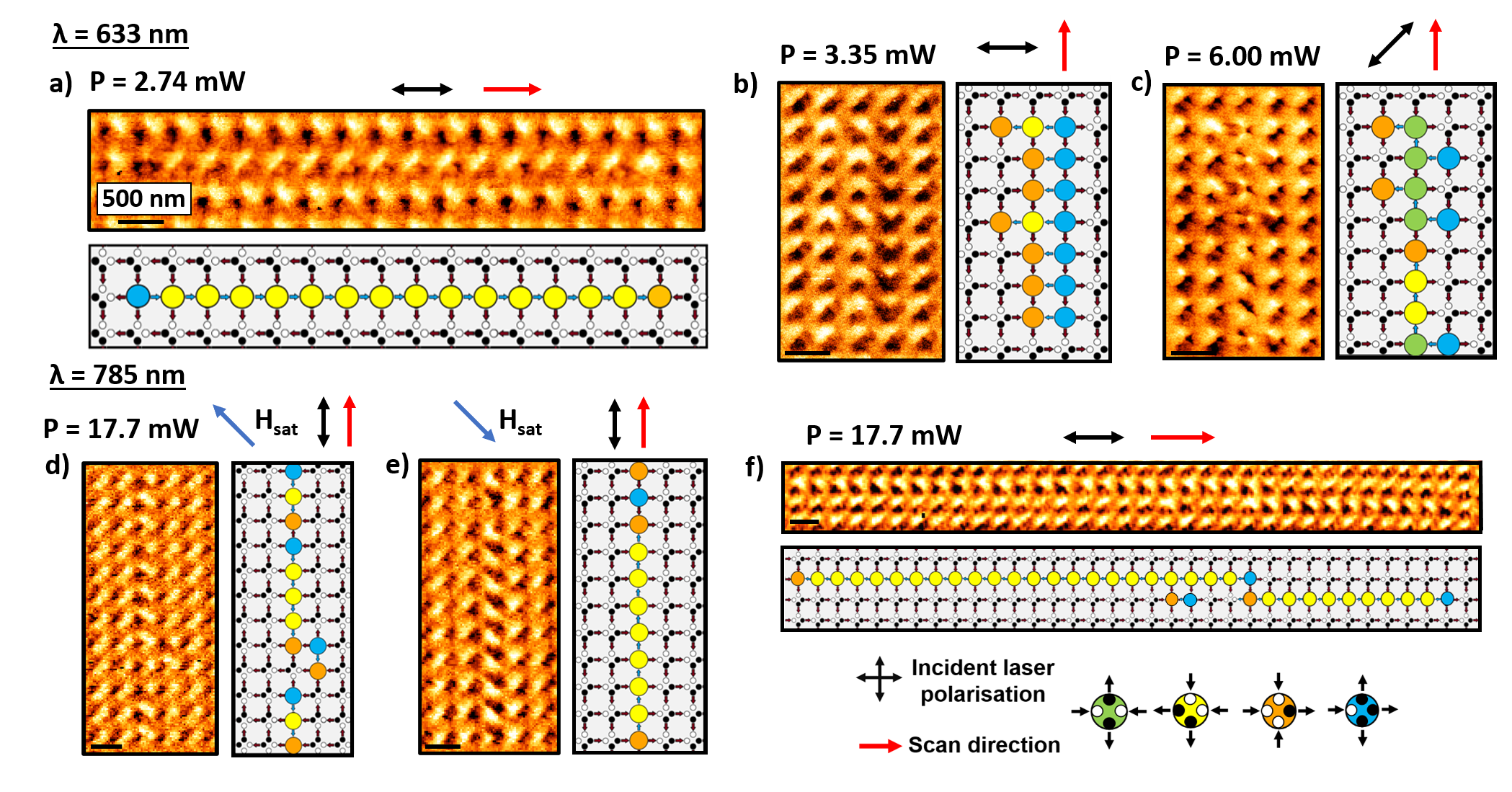}
    \caption{MFM images of switching chains after exposure to $\lambda$ = 633 nm scanning laser with a) polarisation $\parallel$ scan direction writing a chain of type 2 vertices b) polarisation $\perp$ scan direction writing a chain of type 3 vertices and c) polarisation at 45$^\circ$ to the scan direction where both subsets of bars are exposed to the same incident power and both subsets are written. d) and e) show MFM images and schematics of switching via  $\lambda$ = 785 nm after saturating the array in opposite directions. Type 2 chains are written in both cases, ruling out the effect of stray field which would assist switching in one case and oppose switching in the other. f) MFM images and schematics after exposure to a $\lambda$ = 785 nm laser with  polarisation $\parallel$ scan direction. f) shows similar writing characteristics to the shorter wavelength but at $\sim$6$\times$ power increase due to the lower absorption at this wavelength.}
    \label{lines}
\end{figure}

We now explore switching when scanning the laser spot across the array. Here, the beam traverses the entire 30 \textmu m array at 20 \textmu m/s, exposing each nanomagnet for $\sim$ 18 ms. The relative angle between the polarisation and the scan direction is either 0$^\circ$, 45$^\circ$ or 90$^\circ$. Figure \ref{lines}) shows MFM images and corresponding magnetisation schematics of line-scanned writing exposures at laser wavelengths of 633 nm (Figure \ref{lines} a-c)) and 785 nm (Figure \ref{lines} d-f)). At 633 nm wavelength, switching occurs at laser powers as low as 2.74 mW, while 785 nm wavelength switching has an optimum power of 17.7 mW; this is due to the wavelength dependent absorption of the nanomagnetic array (supplementary note 1) and is higher than expected due to wavelength dependent optical losses in the experimental setup (see Methods). For ASI fabricated on Au/SiO$_2$ substrates with higher total absorption, we observe switching across a wider range of nanomagnet dimensions (up to 1 \textmu m bar length) at similar laser powers (supplementary note 5). 

Parallel polarisation and line-scan direction (Figure \ref{lines} a,d-f)) gives rise to long chains of switches where adjacent nanomagnet reversals share a common vertex. Lines written in this configuration comprise a chain of type 2 vertices (yellow circles) with a single type 3 vertex at either end (blue and orange circles). Other than creating the type 3 vertex pair, system energy is not increased as the written type 2 vertices are equal in energy to the initial saturated type 2 background state. Furthermore, the orientation between the scan direction and the nanomagnet means that one end of the bar is always exposed first, maximising the asymmetry in the absorption. Up to 22 consecutive reversals are observed in this configuration. Perpendicular polarisation and scan direction (Figure \ref{lines} b)) results in a type 3 vertex pair at every point on the line-scan (as observed in the static exposures Figure  \ref{dots} c-f)), continually increasing system energy. Up to 7 consecutive reversals are observed in this configuration, demonstrating strong reversal control over highly energetically-unfavourable states. If the laser polarisation is oriented 45$^\circ$ to the scan direction (Figure \ref{lines} c)), both $x,y$-subsets receive equal power and both can switch, resulting in a chain of type 1 vertices (green circles). Here, $\sim$2$\times$ power is required, as expected. 

No magnetic fields are applied during writing and Hall probe measurements reveal stray fields below 1 Oe at the sample. Figure \ref{lines} d,e) show MFM images and vertex type schematics after saturating the same array in opposite initial directions and exposing to a scanning beam ($\lambda$ = 785 nm) with polarisation parallel to the scan direction. Any unintended stray field in the experimental setup that assists reversal in one direction will oppose reversal in the other. Crucially, switching is observed in both cases - ruling out the role of external fields in switching. 

Our results demonstrate deterministic and rewriteable AOMS in isolated and densely-packed nanomagnets. Our methodology requires no external field and operates using simple CW-lasers operating at low-power (2.74 - 6 mW) and low power-densities (1.04 - 2.27 MW/cm$^2$ for 580 nm spot). The absorption profiles allow for attractive optical read/write functionality by tuning the incident wavelength (e.g. write at 633 nm, read at 900 nm).

\subsection*{Reversal mechanism}

\begin{figure}[t!]
    \centering
    \includegraphics[width=\textwidth]{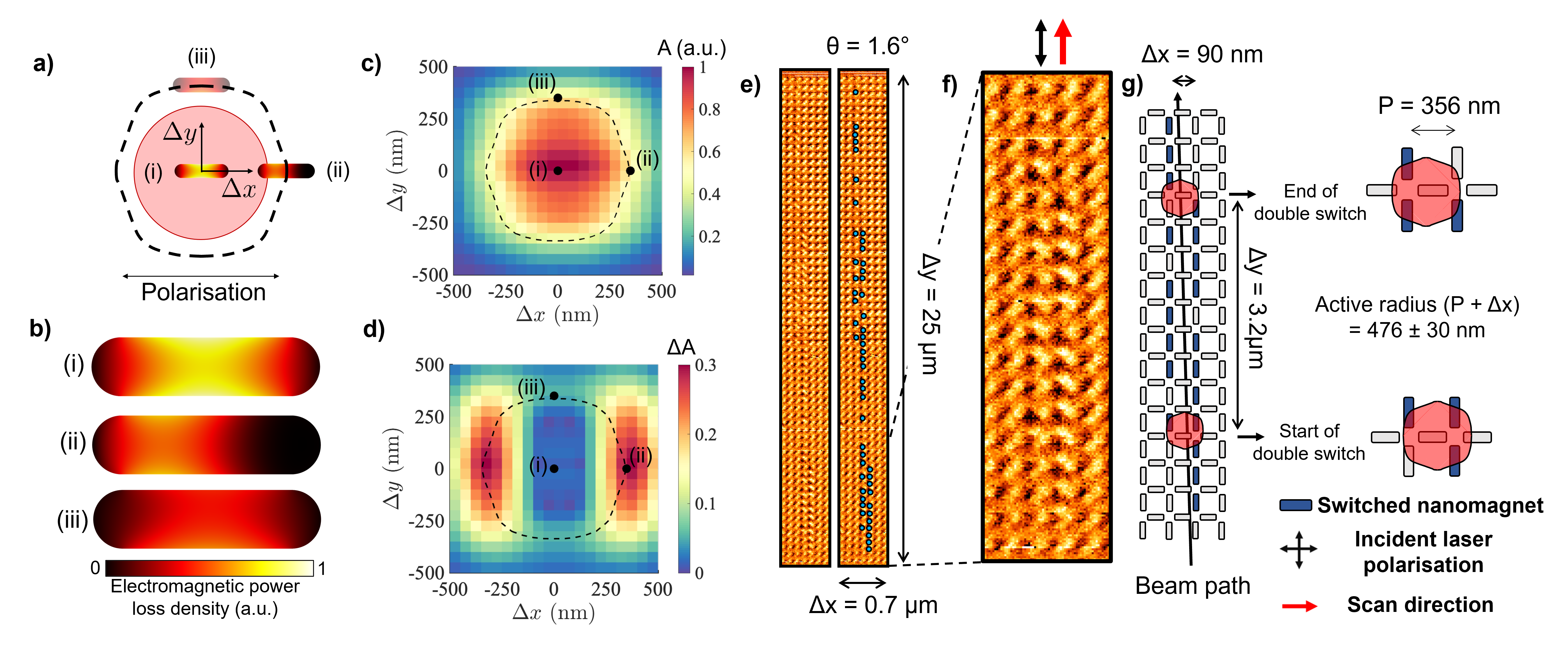}
    \caption{Enhanced asymmetric absorption. a) Simulations schematic whereby a nanomagnet is shifted relative to a central Gaussian beam with a diameter of 580 nm. b) Absorption profiles of the three positions marked in a). Asymmetric absorption profiles are observed when the beam is off-centre. c) Simulated normalised absorption and d) asymmetry factor  $\Delta A(\Delta x, \Delta y)$ for varying shifts, $\Delta$x and $\Delta$y. The dashed shape indicates 50$\%$ total absorption. Nanomagnet dimensions of $L$ = 226 nm, $W$ = 78 nm and $t$ = 20 nm are used.
    e) MFM image (raw and annotated) of a chain of reversals when scanning in $y$ with an offset of 1.6$^{\circ}$. Blue dots correspond to reversed nanomagnets. Two regions of single and double line switches are observed. f) Zoomed in MFM image and g) schematic of region with two columns of switching. At a 1.6$^{\circ}$ angle, the two regions of double switching lasts for 3.2 and 5.2 $\mu$m in the y direction during which the beam moves 90 and 140 nm in x. This gives an active region of 476 $\pm$ 30 nm within which nanomagnets can switch.
    }
    \label{Asymmetric}
\end{figure}

The all-optical switching mechanism requires strong polarisation-dependent absorption of light in the nanomagnets, as well as the existence of an asymmetry to provide a driving force for reversal. The size and shape of the nanomagnets thus have dual roles controlling both magnetic behaviour and the polarisation-dependent optical absorption. Absorption is strongest when polarised along the nanomagnet long-axis, giving selectivity of different bar subsets as observed in Figures \ref{dots} and \ref{lines}. There are three light-absorption mechanisms: the plasmonic antenna resonance; a plasmonic grating effect due to the in-plane nanoarray periodicity; and the interference effect of reflection from the substrate-silica interface. We find the latter mechanism dominates (see supplementary note 3). This is to our advantage since this mechanism allows design freedom over the nanomagnet dimensions and array geometry, as well as greater tolerance to nanofabrication imperfections. Our choice of substrate significantly enhances the total absorption in the nanomagnet up to 16.25 $\times$ compared to conventional Si substrates (see supplementary notes 1).

A key role is played by the location of the laser spot relative to each nanomagnet. We have computed the absorption, $A(\Delta x, \Delta y)$, of light in the Py layer of a nanomagnet in a Gaussian beam where $\Delta x$ and $\Delta y$ are displacements relative to a fixed beam, as shown in Figure \ref{Asymmetric} a). The Gaussian beam has a waist diameter  of $580$ nm at the sample, which matches experimental conditions. We also compute the absorption in four quadrants of the nanomagnet, to evaluate the degree of absorption asymmetry $\Delta A(\Delta x, \Delta y)$ (see Methods). Figure \ref{Asymmetric} b) shows the absorption profiles for three nanomagnet positions. For the central nanomagnet position (i), the absorption is localised in the centre of the nanomagnet. As the nanomagnet is shifted (ii, iii), the absorption profile becomes asymmetric, providing a driving force for magnetic reversal (the implications of this asymmetry are discussed later). Figure \ref{Asymmetric} c) and \ref{Asymmetric} d) show the simulated absorption and absorption asymmetry factor, both normalised to the peak absorption, $A(0,0)$. Asymmetric absorption is a strong effect, with more than $50$ \% of peak absorption occurring at peak asymmetry. This implies that one side of the nanomagnet absorbs $\sim 4\times$ more than the other side. The asymmetric absorption grows for larger displacements, but with diminishing total absorption. Thus the effect of asymmetry is strongest within the illustrated dotted line boundaries of Figure \ref{Asymmetric} c) and d), where the absorption is > 50\%  of the peak value. The shape of the absorption profile (dotted line) is a convolution between the circular beam and bar-shaped nanomagnet. 

When applied to ASI arrays, nanomagnet switching is affected by polarisation and scan direction.  Figure \ref{Asymmetric} e) shows an MFM image, raw (left) and annotated to highlight reversal (right), after scanning a laser along an ASI array. Here the polarisation is aligned in $\hat{y}$. The beam traverses 0.7 $\mu$m and 25 $\mu$m in x and y respectively with an angular offset of 1.6$^{\circ}$. Two regions of double-column switching are observed with lengths of 3.2 $\mu$m and 5.2 $\mu$m in y. Figure \ref{Asymmetric} shows f) an enlarged MFM image and g) schematic of a region where two columns of switches occur (switched nanomagnets are shaded blue). From the angular offset, we deduce that the beam traverses 90 nm and 140 nm in x respectively during these regions. From this we can approximate the diameter D of the active region of the beam using D =  $\Delta x + P$ where P is the period of the ASI array. This gives an active region of D = 476 $\pm$ 30 nm, slightly smaller than the simulated region of 50\% absorption in Figure \ref{Asymmetric} c). As such, when scanning parallel to the polarisation and bar long-axis, a single line of nanomagnets can be switched as D < 2P meaning only one column of bars is illuminated. This is consistently observed throughout Figure \ref{lines}. However, when the scan direction is perpendicular to the polarisation, multiple columns of nanomagnets are seen to switch about 20\% of the time, as shown in Figure \ref{lines} b. We attribute this to the structure of the asymmetric absorption profile in Figure \ref{Asymmetric} d). While the two regions of asymmetry move one after the other for scan direction and polarisation aligned, they move side-by-side for the perpendicular case. The broader reach of the asymmetry in the latter case is consistent with the observation of multiple switched lines. Further evidence of the key role played by the asymmetry is found in the switching fidelity when scanning parallel to the polarisation. In this configuration, one end of the nanomagnet is preferentially excited, maximising asymmetry. We consistently observe higher writing fidelity in this mode of operation. As such, there is strong evidence that the asymmetric absorption provides the driving force for reversal. A result of the asymmetric absorption is the presence of a temperature gradient across the length of the nanomagnet, during the initial stages of excitation, which can facilitate the flow of magnons, spin currents\cite{xiao2010theory} and domain walls\cite{islam2019thermal}. 

The existence of asymmetry in the writing process alone is not sufficient to explain the cause of reversal. There must be an additional microscopic mechanism which is capable of deterministically inverting the magnetisation during each illumination. Here, we begin with a process of deduction. The deterministic switching into any allowed magnetic state with the same sample mounting as illustrated in Figure \ref{lines} d,e) excludes field-driven heat-assisted switching (as in Pancaldi et al\cite{pancaldi2019selective}) from unintentional stray fields in our system. Our deterministic findings are also not consistent with stochastic thermal switching effects from heating nanomagnets beyond the Curie temperature T$_c$ or superparamagnetic limit. The written high-energy, low entropy monopole defect states are never favoured by thermalisation\cite{gartside2018realization}. The observed switching fidelity in Figure \ref{lines} f) of 22 consecutive switches has a corresponding thermalisation probability of 0.5\textsuperscript{22} = \( 2.3 \times 10\textsuperscript{-7} \). We anticipate that the low-power illumination would not heat the lattice temperature above 380 K\cite{pancaldi2019selective}. Magnetometry measurements of a 2 $\times$ 2 mm ASI array with equivalent dimensions reveals only a $\sim$10$\%$ drop in the magnetisation is observed between 120 - 380 K implying that we are not heating close to T$_c$ for these nanomagnet samples (supplementary note 6). 

We observe a single, deterministic reversal for each exposure as opposed to multiple switches across $\sim$ms timescales. IFE and dichroism effects are ruled out, both due to the linearly-polarised exposure and the deterministic inversion functionality we observe. Helicity-dependent effects have a magnetic-field-like symmetry-breaking and so write a specific magnetisation direction, with no effect if the magnetisation is already in that direction - unlike our observed `toggling'. As such, it is likely that the microscopic process involves transient dynamics during the initial stages of excitation when the temperature gradient arising from asymmetric absorption is maximum i.e. as the system is approaching thermodynamic equilibrium. It is plausible that during initial exposure, the electron and spin temperatures at the ends of the nanomagnets rise rapidly, as observed in a number of AOMS studies\cite{ostler2012ultrafast}, leading to rapid demagnetisation. Magnetisation reversal may then proceed via transient spin currents, magnon generation and propagation or domain wall propagation along the absorption-induced temperature gradient. Another possibility is spin redistribution through dynamic changes in the spin bands. There has been recent theoretical and experimental evidence of optically induced spin transfer (OISTR) in ferromagnetic alloys\cite{hofherr2020ultrafast,dewhurst2018laser,willems2020optical} whereby incoming photons may excite electrons between spin minority bands of a heterogeneous material provided that the band structure satisfies necessary requirements. Specifically, there should be an available transition between two spin minority bands at the incident photon energy and no available transitions in the spin majority bands. These studies do not observe switching, yet single-pulse helicity-independent switching is observed in Co/Pt multilayers\cite{vomir2017single} which possess the necessary band structure\cite{uba1996optical}. The phenomena is general to all multi-component magnetic media \cite{willems2020optical} and therefore strong in Ni\textsubscript{50}Fe\textsubscript{50}\cite{hofherr2020ultrafast}, more so than Ni\textsubscript{81}Fe\textsubscript{19}, but not present in Co. 

To test our working hypothesis that the observed AOMS is driven by a plasmonically enhanced OISTR mechanism, we fabricated a set of Ni\textsubscript{50}Fe\textsubscript{50} (20 nm thickness) and Co (8 nm thickness) nanomagnetic arrays patterned on a Au/SiO$_2$ substrate (supplementary note 7), with the prediction that switching fidelity/power threshold would be as good or better in Ni\textsubscript{50}Fe\textsubscript{50} and significantly worse in pure Co. Both predictions were confirmed in the subsequent experiments. For Ni\textsubscript{50}Fe\textsubscript{50}, isolated nanomagnet switches were observed at 3 mW ($\sim$ 14$\%$ reduction in power) across a broader range of nanomagnet dimensions compared to Ni\textsubscript{81}Fe\textsubscript{19}. Conversely, no switches were observed in Co nanostructures up to incident powers of 30 mW, consistent with the OISTR process. Previous OISTR studies focus on ultrafast spin dynamics, and the effects over longer timescales remain unknown. The use of CW exposure, as opposed to ultrafast, suggests OISTR may both stronger and persist for longer than previously expected, highlighting the need for dedicated studies at longer timescales. Whilst OISTR alone can not cause switching, the rapid redistribution of spins during the initial transient stages of excitation may be sufficient to nucleate a reversal.

In summary, we have demonstrated selective and deterministic all-optical magnetic switching of individual and densely-packed ferromagnetic nanostructures using a linearly-polarised, low-power CW laser. We achieve sub-diffraction limit, single nanomagnet switching of diverse ASI microstates including thermally unfavourable high-energy configurations. Our observation of toggle-switching, or deterministic inversion, strongly implies a symmetry-breaking driving force, but is not consistent with a magnetic-field-like force. The optical excitation and absorption within the nanomagnet are strongly asymmetric, yielding a transient temperature gradient and other transient dynamics. This raises fascinating questions about the microscopic reversal mechanism and many avenues for future work. Our results demonstrate high-power ultrafast excitation is not a prerequisite to AOMS, and highlight the need for magneto-optic studies across longer timeframes. The materials employed are cheap and earth-abundant, and we expect switching functionality to be retained across a broad range of substrate and ferromagnetic materials making the technique scaleable and highly compatible with existing technologies. The low-power consumption and cheap cost of the non-specialised CW laser have profound implications across a host of device technologies including data storage, magnonics and non-conventional computing functionalities, particularly neuromorphic and memcomputing hardware. 

\subsection*{Author contributions}
KDS, XX, JCG, RO and WRB conceived the work.\\
KDS drafted the manuscript other than the description of the absorption, with contributions from all authors in editing and revision stages. XX and RO drafted the description of the absorption section.\\
KDS, JCG, AV, OK and HH fabricated the ASI.\\
XX and HH performed the laser illumination protocols.\\
KDS, HH and JCG performed MFM measurements.\\
XX performed FTIR measurements, simulations and absorption calculations. \\
KDS, AV and JCG performed magnetometry measurements.\\
KDS performed simulations of vertex energies.\\
KDS created CGI images.\\

\subsection*{Acknowledgements}
This work was supported by the Leverhulme Trust (RPG-2017-257) to WRB. \\
AV was supported by the EPSRC Centre for Doctoral Training in Advanced Characterisation of Materials (Grant No. EP/L015277/1).\\
Simulations were performed on the Imperial College London Research Computing Service\cite{hpc}.\\
The authors would like to thank Lesley F. Cohen of Imperial College London, Na\"emi Leo of Universidad de Zaragoza and Tom Hayward of The University of Sheffield for enlightening discussion and comments and David Mack for excellent laboratory management.


\subsection*{Data availability statement}
The datasets generated during and/or analysed during the current study are available from the corresponding author on reasonable request.

\subsection*{Simulation details}

The numerical simulations of asymmetric absorption were carried out using the finite element method (FEM) technique (COMSOL Multiphysics). To simulate a Gaussian beam with the diffraction limited spot size, the paraxial Gaussian beam formula was applied along the negative $z$-direction in our simulation. The incidence is polarised in the $x$-direction with its focus point at the structure-substrate interface. In this simulation, a single nanomagnet with dimensions of $L$ = 226 nm, $W$ = 78 nm and $t$ = 20 nm are used. We also include a 20 nm Al$_2$O$_3$ capping on either side of the Py layer to match experimental conditions. The nanomagnet is positioned on the top of a Si / SiO$_2$ (300 nm) substrate with its long axis along the $x$-direction.  In all directions, perfectly matched layer boundary conditions were applied to absorb incident light with minimal reflections. To simulate the normalised absorption and the asymmetric absorption, the nanomagnet is separated into left-upper, left-bottom, right-upper, right-bottom parts and the absorption of the Py layer in these four parts of the nanomagnet corresponding to different displacements, $A_\mathrm{l,u}(\Delta x, \Delta y)$, $A_\mathrm{l,b}(\Delta x, \Delta y)$, $A_\mathrm{r,u}(\Delta x, \Delta y)$, and $A_\mathrm{r,b}(\Delta x, \Delta y)$, is calculated. The absorption of the left and right half sides of the nanomagnet were calculated using $A_\mathrm{l}(\Delta x, \Delta y)=A_\mathrm{l,u}(\Delta x, \Delta y)+A_\mathrm{l,b}(\Delta x, \Delta y)$ and $A_\mathrm{r}(\Delta x, \Delta y)=A_\mathrm{r,u}(\Delta x, \Delta y)+A_\mathrm{r,b}(\Delta x, \Delta y)$. The absorption of the upper and bottom half sides of the nanomagnet were calculated using $A_\mathrm{u}(\Delta x, \Delta y)=A_\mathrm{l,u}(\Delta x, \Delta y)+A_\mathrm{r,u}(\Delta x, \Delta y)$ and $A_\mathrm{b}(\Delta x, \Delta y)=A_\mathrm{l,b}(\Delta x, \Delta y)+A_\mathrm{r,b}(\Delta x, \Delta y)$. The asymmetry was then calculated as $\Delta A(\Delta x, \Delta y)=\sqrt{(A_\mathrm{l}(\Delta x, \Delta y)-A_\mathrm{r}(\Delta x, \Delta y))^2+(A_\mathrm{u}(\Delta x, \Delta y)-A_\mathrm{b}(\Delta x, \Delta y))^2}$. The incident wavelength is 633 nm. The refractive indexes of the Al$_2$O$_3$, Py, SiO$_2$, Si are set 1.7662, 2.3813-3.9415i, 1.4570, 3.8813-0.0189i, respectively.
 
\subsection*{Experimental methods}

Samples were fabricated via electron-beam lithography with liftoff on a Raith eLine system with a bilayer 495K / 950K PMMA resist. Si substrates with a 300 nm SiO\textsubscript{2} layer were purchased commercially. The Au substrates were deposited on an Si/SiO$_{2}$ substrate with the following thicknesses (nm)  Cr(2)/Au(250)/Cr(2)/SiO$_2$(290). Cr is used to aid Au adhesion.  Al\textsubscript{2}O\textsubscript{3}, Ni$_{81}$Fe$_{19}$, Ni$_{50}$Fe$_{50}$ and Co were thermally evaporated at a base pressure of \( 2 \times 10 \textsuperscript{-6}\) mbar.  For samples fabricated on Si / SiO\textsubscript{2}, a 15 nm Al$_2$O$_3$ layer is deposited on either side of the nanomagnet to protect against oxidation. For samples fabricated on Au / SiO$_2$ substrates, a 4 nm Al$_2$O$_3$ coating on the top layer is used to protect oxidation. The presence and thickness of Al$_2$O$_3$ does not significantly affect light absorption. 

Magnetic force micrographs were produced on a Dimension 3100 using commercially available low-moment and normal-moment MFM tips.
\subsection*{Illumination experiments}

Laser illumination by continuous wave lasers with different wavelengths (633 nm and 785 nm) were focused to a diffraction limited spot on the sample through a confocal Raman microscope (alpha300 RSA+, WITec). The light was linearly polarized and focused by a 100$\times$ (NA = 0.9, Zeiss) microscope objective. The focal spot was scanned to illuminate the locations of interest. To achieve the fastest line scanning speed, the instrument was simply scanned between two pre-defined points. For slower scans, 50 points were defined along each scan line and the detector integration time was used to control the dwell time. It should be noted that the integration time is usually used to control spectra collection, but here this parameter controls the illumination time. The power of the beam was measured by a power meter (PM100D, Thorlabs) attached to the objective turret and all powers stated are measured before focusing. The power transmitted through the objective is $\leq$ 90 $\%$, and may be lower due to optical losses in the set up. Measurements suggest that $\sim$ 65 \% and  $\sim$ 45 \% of the power may be reaching the sample at wavelengths of 633 nm and 785 nm respectively.

\label{Bibliography}
\bibliography{bib.bib}

\begin{thebibliography}{10}
\urlstyle{rm}
\expandafter\ifx\csname url\endcsname\relax
  \def\url#1{\texttt{#1}}\fi
\expandafter\ifx\csname urlprefix\endcsname\relax\def\urlprefix{URL }\fi
\expandafter\ifx\csname doiprefix\endcsname\relax\def\doiprefix{DOI: }\fi
\providecommand{\bibinfo}[2]{#2}
\providecommand{\eprint}[2][]{\url{#2}}

\bibitem{kovalenko1986photoinduced}
\bibinfo{author}{Kovalenko, V.} \& \bibinfo{author}{Nagaev, {\'E}.~L.}
\newblock \bibinfo{journal}{\bibinfo{title}{Photoinduced magnetism}}.
\newblock {\emph{\JournalTitle{Soviet Physics Uspekhi}}}
  \textbf{\bibinfo{volume}{29}}, \bibinfo{pages}{297} (\bibinfo{year}{1986}).

\bibitem{pancaldi2019selective}
\bibinfo{author}{Pancaldi, M.}, \bibinfo{author}{Leo, N.} \&
  \bibinfo{author}{Vavassori, P.}
\newblock \bibinfo{journal}{\bibinfo{title}{Selective and fast plasmon-assisted
  photo-heating of nanomagnets}}.
\newblock {\emph{\JournalTitle{Nanoscale}}} \textbf{\bibinfo{volume}{11}},
  \bibinfo{pages}{7656--7666} (\bibinfo{year}{2019}).

\bibitem{kryder2008heat}
\bibinfo{author}{Kryder, M.~H.} \emph{et~al.}
\newblock \bibinfo{journal}{\bibinfo{title}{Heat assisted magnetic recording}}.
\newblock {\emph{\JournalTitle{Proceedings of the IEEE}}}
  \textbf{\bibinfo{volume}{96}}, \bibinfo{pages}{1810--1835}
  (\bibinfo{year}{2008}).

\bibitem{stanciu2007all}
\bibinfo{author}{Stanciu, C.~D.} \emph{et~al.}
\newblock \bibinfo{journal}{\bibinfo{title}{All-optical magnetic recording with
  circularly polarized light}}.
\newblock {\emph{\JournalTitle{Physical review letters}}}
  \textbf{\bibinfo{volume}{99}}, \bibinfo{pages}{047601}
  (\bibinfo{year}{2007}).

\bibitem{kirilyuk2010ultrafast}
\bibinfo{author}{Kirilyuk, A.}, \bibinfo{author}{Kimel, A.~V.} \&
  \bibinfo{author}{Rasing, T.}
\newblock \bibinfo{journal}{\bibinfo{title}{Ultrafast optical manipulation of
  magnetic order}}.
\newblock {\emph{\JournalTitle{Reviews of Modern Physics}}}
  \textbf{\bibinfo{volume}{82}}, \bibinfo{pages}{2731} (\bibinfo{year}{2010}).

\bibitem{kimel2019writing}
\bibinfo{author}{Kimel, A.~V.} \& \bibinfo{author}{Li, M.}
\newblock \bibinfo{journal}{\bibinfo{title}{Writing magnetic memory with
  ultrashort light pulses}}.
\newblock {\emph{\JournalTitle{Nature Reviews Materials}}}
  \textbf{\bibinfo{volume}{4}}, \bibinfo{pages}{189--200}
  (\bibinfo{year}{2019}).

\bibitem{el2016two}
\bibinfo{author}{El~Hadri, M.~S.} \emph{et~al.}
\newblock \bibinfo{journal}{\bibinfo{title}{Two types of all-optical
  magnetization switching mechanisms using femtosecond laser pulses}}.
\newblock {\emph{\JournalTitle{Physical review B}}}
  \textbf{\bibinfo{volume}{94}}, \bibinfo{pages}{064412}
  (\bibinfo{year}{2016}).

\bibitem{medapalli2017multiscale}
\bibinfo{author}{Medapalli, R.} \emph{et~al.}
\newblock \bibinfo{journal}{\bibinfo{title}{Multiscale dynamics of
  helicity-dependent all-optical magnetization reversal in ferromagnetic co/pt
  multilayers}}.
\newblock {\emph{\JournalTitle{Physical review B}}}
  \textbf{\bibinfo{volume}{96}}, \bibinfo{pages}{224421}
  (\bibinfo{year}{2017}).

\bibitem{ellis2016all}
\bibinfo{author}{Ellis, M.~O.}, \bibinfo{author}{Fullerton, E.~E.} \&
  \bibinfo{author}{Chantrell, R.~W.}
\newblock \bibinfo{journal}{\bibinfo{title}{All-optical switching in granular
  ferromagnets caused by magnetic circular dichroism}}.
\newblock {\emph{\JournalTitle{Scientific reports}}}
  \textbf{\bibinfo{volume}{6}}, \bibinfo{pages}{1--9} (\bibinfo{year}{2016}).

\bibitem{lambert2014all}
\bibinfo{author}{Lambert, C.-H.} \emph{et~al.}
\newblock \bibinfo{journal}{\bibinfo{title}{All-optical control of
  ferromagnetic thin films and nanostructures}}.
\newblock {\emph{\JournalTitle{Science}}} \textbf{\bibinfo{volume}{345}},
  \bibinfo{pages}{1337--1340} (\bibinfo{year}{2014}).

\bibitem{kichin2019multiple}
\bibinfo{author}{Kichin, G.} \emph{et~al.}
\newblock \bibinfo{journal}{\bibinfo{title}{From multiple-to single-pulse
  all-optical helicity-dependent switching in ferromagnetic co/pt
  multilayers}}.
\newblock {\emph{\JournalTitle{Physical Review Applied}}}
  \textbf{\bibinfo{volume}{12}}, \bibinfo{pages}{024019}
  (\bibinfo{year}{2019}).

\bibitem{ostler2012ultrafast}
\bibinfo{author}{Ostler, T.} \emph{et~al.}
\newblock \bibinfo{journal}{\bibinfo{title}{Ultrafast heating as a sufficient
  stimulus for magnetization reversal in a ferrimagnet}}.
\newblock {\emph{\JournalTitle{Nature communications}}}
  \textbf{\bibinfo{volume}{3}}, \bibinfo{pages}{1--6} (\bibinfo{year}{2012}).

\bibitem{stanciu2007subpicosecond}
\bibinfo{author}{Stanciu, C.} \emph{et~al.}
\newblock \bibinfo{journal}{\bibinfo{title}{Subpicosecond magnetization
  reversal across ferrimagnetic compensation points}}.
\newblock {\emph{\JournalTitle{Physical review letters}}}
  \textbf{\bibinfo{volume}{99}}, \bibinfo{pages}{217204}
  (\bibinfo{year}{2007}).

\bibitem{radu2011transient}
\bibinfo{author}{Radu, I.} \emph{et~al.}
\newblock \bibinfo{journal}{\bibinfo{title}{Transient ferromagnetic-like state
  mediating ultrafast reversal of antiferromagnetically coupled spins}}.
\newblock {\emph{\JournalTitle{Nature}}} \textbf{\bibinfo{volume}{472}},
  \bibinfo{pages}{205--208} (\bibinfo{year}{2011}).

\bibitem{el2017materials}
\bibinfo{author}{El~Hadri, M.~S.}, \bibinfo{author}{Hehn, M.},
  \bibinfo{author}{Malinowski, G.} \& \bibinfo{author}{Mangin, S.}
\newblock \bibinfo{journal}{\bibinfo{title}{Materials and devices for
  all-optical helicity-dependent switching}}.
\newblock {\emph{\JournalTitle{Journal of Physics D: Applied Physics}}}
  \textbf{\bibinfo{volume}{50}}, \bibinfo{pages}{133002}
  (\bibinfo{year}{2017}).

\bibitem{le2012demonstration}
\bibinfo{author}{Le~Guyader, L.} \emph{et~al.}
\newblock \bibinfo{journal}{\bibinfo{title}{Demonstration of laser induced
  magnetization reversal in gdfeco nanostructures}}.
\newblock {\emph{\JournalTitle{Applied Physics Letters}}}
  \textbf{\bibinfo{volume}{101}}, \bibinfo{pages}{022410}
  (\bibinfo{year}{2012}).

\bibitem{beaurepaire1996ultrafast}
\bibinfo{author}{Beaurepaire, E.}, \bibinfo{author}{Merle, J.-C.},
  \bibinfo{author}{Daunois, A.} \& \bibinfo{author}{Bigot, J.-Y.}
\newblock \bibinfo{journal}{\bibinfo{title}{Ultrafast spin dynamics in
  ferromagnetic nickel}}.
\newblock {\emph{\JournalTitle{Physical review letters}}}
  \textbf{\bibinfo{volume}{76}}, \bibinfo{pages}{4250} (\bibinfo{year}{1996}).

\bibitem{xu2017ultrafast}
\bibinfo{author}{Xu, Y.} \emph{et~al.}
\newblock \bibinfo{journal}{\bibinfo{title}{Ultrafast magnetization
  manipulation using single femtosecond light and hot-electron pulses}}.
\newblock {\emph{\JournalTitle{Advanced Materials}}}
  \textbf{\bibinfo{volume}{29}}, \bibinfo{pages}{1703474}
  (\bibinfo{year}{2017}).

\bibitem{banerjee2020single}
\bibinfo{author}{Banerjee, C.} \emph{et~al.}
\newblock \bibinfo{journal}{\bibinfo{title}{Single pulse all-optical toggle
  switching of magnetization without gadolinium in the ferrimagnet mn2ruxga}}.
\newblock {\emph{\JournalTitle{Nature communications}}}
  \textbf{\bibinfo{volume}{11}}, \bibinfo{pages}{1--6} (\bibinfo{year}{2020}).

\bibitem{gorchon2017single}
\bibinfo{author}{Gorchon, J.} \emph{et~al.}
\newblock \bibinfo{journal}{\bibinfo{title}{Single shot ultrafast all optical
  magnetization switching of ferromagnetic co/pt multilayers}}.
\newblock {\emph{\JournalTitle{Applied physics letters}}}
  \textbf{\bibinfo{volume}{111}}, \bibinfo{pages}{042401}
  (\bibinfo{year}{2017}).

\bibitem{iihama2018single}
\bibinfo{author}{Iihama, S.} \emph{et~al.}
\newblock \bibinfo{journal}{\bibinfo{title}{Single-shot multi-level all-optical
  magnetization switching mediated by spin transport}}.
\newblock {\emph{\JournalTitle{Advanced Materials}}}
  \textbf{\bibinfo{volume}{30}}, \bibinfo{pages}{1804004}
  (\bibinfo{year}{2018}).

\bibitem{igarashi2020engineering}
\bibinfo{author}{Igarashi, J.} \emph{et~al.}
\newblock \bibinfo{journal}{\bibinfo{title}{Engineering single-shot all-optical
  switching of ferromagnetic materials}}.
\newblock {\emph{\JournalTitle{Nano Letters}}} \textbf{\bibinfo{volume}{20}},
  \bibinfo{pages}{8654--8660} (\bibinfo{year}{2020}).

\bibitem{vomir2017single}
\bibinfo{author}{Vomir, M.}, \bibinfo{author}{Albrecht, M.} \&
  \bibinfo{author}{Bigot, J.-Y.}
\newblock \bibinfo{journal}{\bibinfo{title}{Single shot all optical switching
  of intrinsic micron size magnetic domains of a pt/co/pt ferromagnetic
  stack}}.
\newblock {\emph{\JournalTitle{Applied Physics Letters}}}
  \textbf{\bibinfo{volume}{111}}, \bibinfo{pages}{242404}
  (\bibinfo{year}{2017}).

\bibitem{camsari2017stochastic}
\bibinfo{author}{Camsari, K.~Y.}, \bibinfo{author}{Faria, R.},
  \bibinfo{author}{Sutton, B.~M.} \& \bibinfo{author}{Datta, S.}
\newblock \bibinfo{journal}{\bibinfo{title}{Stochastic p-bits for invertible
  logic}}.
\newblock {\emph{\JournalTitle{Physical Review X}}}
  \textbf{\bibinfo{volume}{7}}, \bibinfo{pages}{031014} (\bibinfo{year}{2017}).

\bibitem{caravelli2020logical}
\bibinfo{author}{Caravelli, F.} \& \bibinfo{author}{Nisoli, C.}
\newblock \bibinfo{journal}{\bibinfo{title}{Logical gates embedding in
  artificial spin ice}}.
\newblock {\emph{\JournalTitle{New Journal of Physics}}}
  \textbf{\bibinfo{volume}{22}}, \bibinfo{pages}{103052}
  (\bibinfo{year}{2020}).

\bibitem{tanaka2019recent}
\bibinfo{author}{Tanaka, G.} \emph{et~al.}
\newblock \bibinfo{journal}{\bibinfo{title}{Recent advances in physical
  reservoir computing: A review}}.
\newblock {\emph{\JournalTitle{Neural Networks}}}
  \textbf{\bibinfo{volume}{115}}, \bibinfo{pages}{100--123}
  (\bibinfo{year}{2019}).

\bibitem{markovic2020physics}
\bibinfo{author}{Markovi{\'c}, D.}, \bibinfo{author}{Mizrahi, A.},
  \bibinfo{author}{Querlioz, D.} \& \bibinfo{author}{Grollier, J.}
\newblock \bibinfo{journal}{\bibinfo{title}{Physics for neuromorphic
  computing}}.
\newblock {\emph{\JournalTitle{Nature Reviews Physics}}}
  \textbf{\bibinfo{volume}{2}}, \bibinfo{pages}{499--510}
  (\bibinfo{year}{2020}).

\bibitem{gartside2021reservoir}
\bibinfo{author}{Gartside, J.~C.} \emph{et~al.}
\newblock \bibinfo{journal}{\bibinfo{title}{Reconfigurable training, reservoir
  computing and spin-wave fingerprinting in an artificial spin-vortex ice}}.
\newblock {\emph{\JournalTitle{arXiv preprint arXiv:2107.08941}}}
  (\bibinfo{year}{2021}).

\bibitem{jensen2018computation}
\bibinfo{author}{Jensen, J.~H.}, \bibinfo{author}{Folven, E.} \&
  \bibinfo{author}{Tufte, G.}
\newblock \bibinfo{title}{Computation in artificial spin ice}.
\newblock In \emph{\bibinfo{booktitle}{Artificial Life Conference
  Proceedings}}, \bibinfo{pages}{15--22} (\bibinfo{organization}{MIT Press},
  \bibinfo{year}{2018}).

\bibitem{jensen2020reservoir}
\bibinfo{author}{Jensen, J.~H.} \& \bibinfo{author}{Tufte, G.}
\newblock \bibinfo{title}{Reservoir computing in artificial spin ice}.
\newblock In \emph{\bibinfo{booktitle}{Artificial Life Conference
  Proceedings}}, \bibinfo{pages}{376--383} (\bibinfo{organization}{MIT Press},
  \bibinfo{year}{2020}).

\bibitem{hon2021numerical}
\bibinfo{author}{Hon, K.} \emph{et~al.}
\newblock \bibinfo{journal}{\bibinfo{title}{Numerical simulation of artificial
  spin ice for reservoir computing}}.
\newblock {\emph{\JournalTitle{Applied Physics Express}}}
  \textbf{\bibinfo{volume}{14}}, \bibinfo{pages}{033001}
  (\bibinfo{year}{2021}).

\bibitem{lendinez2019magnetization}
\bibinfo{author}{Lendinez, S.} \& \bibinfo{author}{Jungfleisch, M.}
\newblock \bibinfo{journal}{\bibinfo{title}{Magnetization dynamics in
  artificial spin ice}}.
\newblock {\emph{\JournalTitle{Journal of Physics: Condensed Matter}}}
  \textbf{\bibinfo{volume}{32}}, \bibinfo{pages}{013001}
  (\bibinfo{year}{2019}).

\bibitem{chumak2017magnonic}
\bibinfo{author}{Chumak, A.}, \bibinfo{author}{Serga, A.} \&
  \bibinfo{author}{Hillebrands, B.}
\newblock \bibinfo{journal}{\bibinfo{title}{Magnonic crystals for data
  processing}}.
\newblock {\emph{\JournalTitle{Journal of Physics D: Applied Physics}}}
  \textbf{\bibinfo{volume}{50}}, \bibinfo{pages}{244001}
  (\bibinfo{year}{2017}).

\bibitem{barman2020magnetization}
\bibinfo{author}{Barman, A.}, \bibinfo{author}{Mondal, S.},
  \bibinfo{author}{Sahoo, S.} \& \bibinfo{author}{De, A.}
\newblock \bibinfo{journal}{\bibinfo{title}{Magnetization dynamics of nanoscale
  magnetic materials: A perspective}}.
\newblock {\emph{\JournalTitle{Journal of Applied Physics}}}
  \textbf{\bibinfo{volume}{128}}, \bibinfo{pages}{170901}
  (\bibinfo{year}{2020}).

\bibitem{kaffash2021nanomagnonics}
\bibinfo{author}{Kaffash, M.~T.}, \bibinfo{author}{Lendinez, S.} \&
  \bibinfo{author}{Jungfleisch, M.~B.}
\newblock \bibinfo{journal}{\bibinfo{title}{Nanomagnonics with artificial spin
  ice}}.
\newblock {\emph{\JournalTitle{Physics Letters A}}}
  \textbf{\bibinfo{volume}{402}}, \bibinfo{pages}{127364}
  (\bibinfo{year}{2021}).

\bibitem{gartside2021reconfigurable}
\bibinfo{author}{Gartside, J.~C.} \emph{et~al.}
\newblock \bibinfo{journal}{\bibinfo{title}{Reconfigurable magnonic
  mode-hybridisation and spectral control in a bicomponent artificial spin
  ice}}.
\newblock {\emph{\JournalTitle{Nature Communications}}}
  \textbf{\bibinfo{volume}{12}}, \bibinfo{pages}{1--9} (\bibinfo{year}{2021}).

\bibitem{stenning2020magnonic}
\bibinfo{author}{Stenning, K.~D.} \emph{et~al.}
\newblock \bibinfo{journal}{\bibinfo{title}{Magnonic bending, phase shifting
  and interferometry in a 2d reconfigurable nanodisk crystal}}.
\newblock {\emph{\JournalTitle{ACS nano}}}  (\bibinfo{year}{2020}).

\bibitem{gartside2020current}
\bibinfo{author}{Gartside, J.~C.} \emph{et~al.}
\newblock \bibinfo{journal}{\bibinfo{title}{Current-controlled nanomagnetic
  writing for reconfigurable magnonic crystals}}.
\newblock {\emph{\JournalTitle{Communications Physics}}}
  \textbf{\bibinfo{volume}{3}}, \bibinfo{pages}{1--8} (\bibinfo{year}{2020}).

\bibitem{dion2022observation}
\bibinfo{author}{Dion, T.} \emph{et~al.}
\newblock \bibinfo{journal}{\bibinfo{title}{Observation and control of
  collective spin-wave mode hybridization in chevron arrays and in square,
  staircase, and brickwork artificial spin ices}}.
\newblock {\emph{\JournalTitle{Physical Review Research}}}
  \textbf{\bibinfo{volume}{4}}, \bibinfo{pages}{013107} (\bibinfo{year}{2022}).

\bibitem{vanstone2021spectral}
\bibinfo{author}{Vanstone, A.} \emph{et~al.}
\newblock \bibinfo{journal}{\bibinfo{title}{Spectral-fingerprinting: Microstate
  readout via remanence ferromagnetic resonance in artificial spin systems}}.
\newblock {\emph{\JournalTitle{arXiv preprint arXiv:2106.04406}}}
  (\bibinfo{year}{2021}).

\bibitem{gypens2021thermoplasmonic}
\bibinfo{author}{Gypens, P.}, \bibinfo{author}{Leo, N.},
  \bibinfo{author}{Menniti, M.}, \bibinfo{author}{Vavassori, P.} \&
  \bibinfo{author}{Leliaert, J.}
\newblock \bibinfo{journal}{\bibinfo{title}{Thermoplasmonic nanomagnetic logic
  gates}}.
\newblock {\emph{\JournalTitle{arXiv preprint arXiv:2110.14212}}}
  (\bibinfo{year}{2021}).

\bibitem{wang2016rewritable}
\bibinfo{author}{Wang, Y.-L.} \emph{et~al.}
\newblock \bibinfo{journal}{\bibinfo{title}{Rewritable artificial magnetic
  charge ice}}.
\newblock {\emph{\JournalTitle{Science}}} \textbf{\bibinfo{volume}{352}},
  \bibinfo{pages}{962--966} (\bibinfo{year}{2016}).

\bibitem{gartside2016novel}
\bibinfo{author}{Gartside, J.}, \bibinfo{author}{Burn, D.},
  \bibinfo{author}{Cohen, L.} \& \bibinfo{author}{Branford, W.}
\newblock \bibinfo{journal}{\bibinfo{title}{A novel method for the injection
  and manipulation of magnetic charge states in nanostructures}}.
\newblock {\emph{\JournalTitle{Scientific reports}}}
  \textbf{\bibinfo{volume}{6}}, \bibinfo{pages}{32864} (\bibinfo{year}{2016}).

\bibitem{gartside2018realization}
\bibinfo{author}{Gartside, J.~C.} \emph{et~al.}
\newblock \bibinfo{journal}{\bibinfo{title}{Realization of ground state in
  artificial kagome spin ice via topological defect-driven magnetic writing}}.
\newblock {\emph{\JournalTitle{Nature nanotechnology}}}
  \textbf{\bibinfo{volume}{13}}, \bibinfo{pages}{53} (\bibinfo{year}{2018}).

\bibitem{wang2006artificial}
\bibinfo{author}{Wang, .~R.} \emph{et~al.}
\newblock \bibinfo{journal}{\bibinfo{title}{Artificial ‘spin ice’in a
  geometrically frustrated lattice of nanoscale ferromagnetic islands}}.
\newblock {\emph{\JournalTitle{Nature}}} \textbf{\bibinfo{volume}{439}},
  \bibinfo{pages}{303--306} (\bibinfo{year}{2006}).

\bibitem{skjaervo2020advances}
\bibinfo{author}{Skj{\ae}rv{\o}, S.~H.}, \bibinfo{author}{Marrows, C.~H.},
  \bibinfo{author}{Stamps, R.~L.} \& \bibinfo{author}{Heyderman, L.~J.}
\newblock \bibinfo{journal}{\bibinfo{title}{Advances in artificial spin ice}}.
\newblock {\emph{\JournalTitle{Nature Reviews Physics}}}
  \textbf{\bibinfo{volume}{2}}, \bibinfo{pages}{13--28} (\bibinfo{year}{2020}).

\bibitem{xiao2010theory}
\bibinfo{author}{Xiao, J.} \emph{et~al.}
\newblock \bibinfo{journal}{\bibinfo{title}{Theory of magnon-driven spin
  seebeck effect}}.
\newblock {\emph{\JournalTitle{Physical Review B}}}
  \textbf{\bibinfo{volume}{81}}, \bibinfo{pages}{214418}
  (\bibinfo{year}{2010}).

\bibitem{islam2019thermal}
\bibinfo{author}{Islam, M.~T.}, \bibinfo{author}{Wang, X.} \&
  \bibinfo{author}{Wang, X.}
\newblock \bibinfo{journal}{\bibinfo{title}{Thermal gradient driven domain wall
  dynamics}}.
\newblock {\emph{\JournalTitle{Journal of Physics: Condensed Matter}}}
  \textbf{\bibinfo{volume}{31}}, \bibinfo{pages}{455701}
  (\bibinfo{year}{2019}).

\bibitem{hofherr2020ultrafast}
\bibinfo{author}{Hofherr, M.} \emph{et~al.}
\newblock \bibinfo{journal}{\bibinfo{title}{Ultrafast optically induced spin
  transfer in ferromagnetic alloys}}.
\newblock {\emph{\JournalTitle{Science advances}}}
  \textbf{\bibinfo{volume}{6}}, \bibinfo{pages}{eaay8717}
  (\bibinfo{year}{2020}).

\bibitem{dewhurst2018laser}
\bibinfo{author}{Dewhurst, J.~K.}, \bibinfo{author}{Elliott, P.},
  \bibinfo{author}{Shallcross, S.}, \bibinfo{author}{Gross, E.~K.} \&
  \bibinfo{author}{Sharma, S.}
\newblock \bibinfo{journal}{\bibinfo{title}{Laser-induced intersite spin
  transfer}}.
\newblock {\emph{\JournalTitle{Nano letters}}} \textbf{\bibinfo{volume}{18}},
  \bibinfo{pages}{1842--1848} (\bibinfo{year}{2018}).

\bibitem{willems2020optical}
\bibinfo{author}{Willems, F.} \emph{et~al.}
\newblock \bibinfo{journal}{\bibinfo{title}{Optical inter-site spin transfer
  probed by energy and spin-resolved transient absorption spectroscopy}}.
\newblock {\emph{\JournalTitle{Nature communications}}}
  \textbf{\bibinfo{volume}{11}}, \bibinfo{pages}{1--7} (\bibinfo{year}{2020}).

\bibitem{uba1996optical}
\bibinfo{author}{Uba, S.} \emph{et~al.}
\newblock \bibinfo{journal}{\bibinfo{title}{Optical and magneto-optical
  properties of co/pt multilayers}}.
\newblock {\emph{\JournalTitle{Physical Review B}}}
  \textbf{\bibinfo{volume}{53}}, \bibinfo{pages}{6526} (\bibinfo{year}{1996}).

\bibitem{hpc}
\bibinfo{title}{Imperial college research computing service}.
\newblock \bibinfo{note}{DOI: 10.14469/hpc/2232}.

\bibitem{edward1985handbook}
\bibinfo{author}{Edward, D.~P.} \& \bibinfo{author}{Palik, I.}
\newblock \bibinfo{title}{Handbook of optical constants of solids}
  (\bibinfo{year}{1985}).

\bibitem{johnson1972optical}
\bibinfo{author}{Johnson, P.~B.} \& \bibinfo{author}{Christy, R.-W.}
\newblock \bibinfo{journal}{\bibinfo{title}{Optical constants of the noble
  metals}}.
\newblock {\emph{\JournalTitle{Physical review B}}}
  \textbf{\bibinfo{volume}{6}}, \bibinfo{pages}{4370} (\bibinfo{year}{1972}).

\bibitem{tikuivsis2017optical}
\bibinfo{author}{Tikui{\v{s}}is, K.~K.} \emph{et~al.}
\newblock \bibinfo{journal}{\bibinfo{title}{Optical and magneto-optical
  properties of permalloy thin films in 0.7--6.4 ev photon energy range}}.
\newblock {\emph{\JournalTitle{Materials \& Design}}}
  \textbf{\bibinfo{volume}{114}}, \bibinfo{pages}{31--39}
  (\bibinfo{year}{2017}).

\bibitem{kuz2005shape}
\bibinfo{author}{Kuz’min, M.}
\newblock \bibinfo{journal}{\bibinfo{title}{Shape of temperature dependence of
  spontaneous magnetization of ferromagnets: quantitative analysis}}.
\newblock {\emph{\JournalTitle{Physical review letters}}}
  \textbf{\bibinfo{volume}{94}}, \bibinfo{pages}{107204}
  (\bibinfo{year}{2005}).

\end{thebibliography}

\subsection*{Supplementary Information}

\subsection*{Supplementary note 1 - Enhanced total absorption via substrate design}
\begin{figure*}[htbp]
    \centering
    \includegraphics[width=\textwidth]{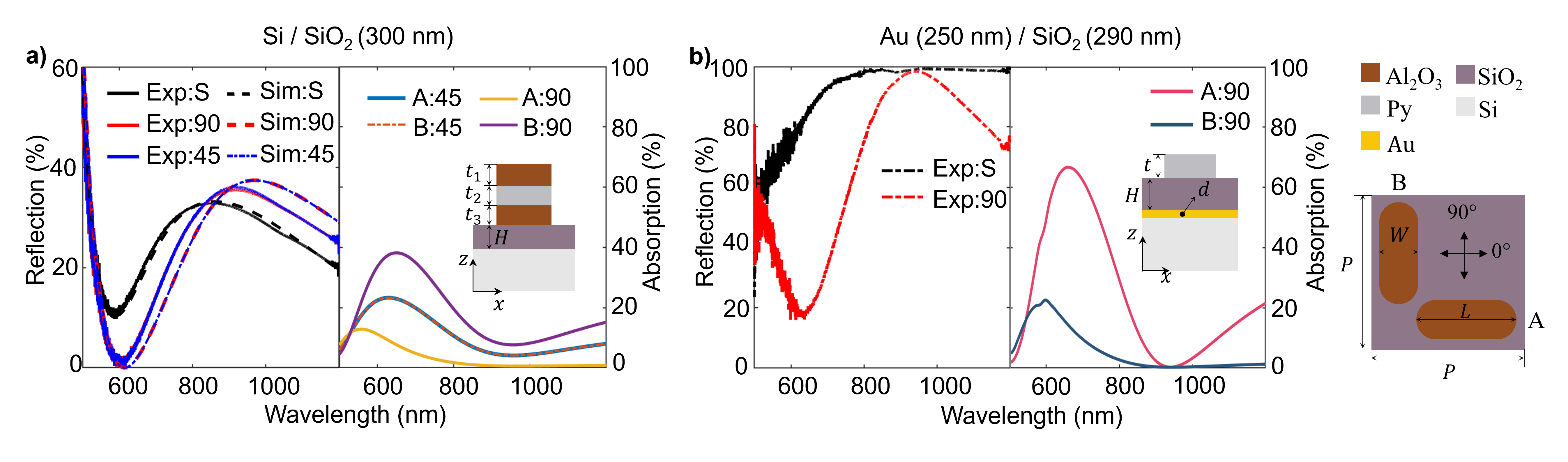}
    \caption*{\textbf{Supporting Figure S1.}  Experimental (Exp) and simulated (Sim) reflection spectra and calculated absorption of a 2 mm$\times$2 mm nanomagnet array (90$^\circ$ / 45$^\circ$ incident polarisation) fabricated on a) Si / SiO$_2$ (300 nm) and b) Au (250 nm) / SiO$_2$ (290 nm) substrates (S). This system significantly reduces reflections and thereby enhances absorption in the nanomagnets.
    In a) sample dimensions are $t_1$ = 15 nm, $t_2$ = 20 nm, $t_3$ = 15 nm, $H$ = 300 nm. Each Py nanomagnet is enclosed in two Al$_2$O$_3$ coatings to protect against oxidation. In b) sample dimensions are $d$ = 250 nm, $H$ = 290 nm, $t$ = 20 nm. Here, a 4nm Al$_2$O$_3$ cap is used. Here we simulate a square ASI nanomagnet array with dimensions of $L$ = 226 nm, $W$ = 78 nm and $P$ = 356 nm. Absorption in the horizontal (A) and vertical (B) Py bars are shown for laser polarisation angles of 45$^\circ$ and 90$^\circ$ relative to the horizontal, highlighting the contrast and subset selectivity afforded. Up to 65$\%$ and 38$\%$ absorption is achieved in the Au/SiO$_2$ and Si/SiO$_2$ respectively.}
    \label{Si_absorption2}
\end{figure*}
Supporting Figure S1 a) compares experimental (`Exp') and simulated (`Sim') Fourier Transform Infrared (FTIR) reflection spectra for the Si/SiO$_2$ substrates with and without nanomagnet arrays. Here, Py nanomagnets are enclosed by two Al$_2$O$_3$ layers (15 nm) to protect against oxidation. Excellent experimental-simulation correspondence is observed. SEM images are provided in supplementary note 2. The addition of the Py nanomagnet layer introduces absorption and reflection that modifies the partial reflections from the substrate-silica and silica-air interfaces. Strong absorption in the nanomagnets corresponds to the condition where these partial reflections destructively interfere, as shown by the reflectivity minimum near 600 nm in Supporting Figure S1 a). Remarkably for $\sim$ 28\% fill factor of the Py ASI system, the total reflection from the device can be $<$ 1\%, leading to a calculated absorption of $38$\%, as shown in Supporting Figure S1 a). Furthermore, the high nanomagnet aspect-ratio provides an optical polarisation response; light is absorbed dominantly in antennas whose long axis is aligned to the optical polarisation. Supporting Figure S1 a) shows that at a wavelength of 633 nm, the absorption ratio for polarisation along long and short axes is a factor $>4$. This enables selective optical switching via incident polarisation control. Supporting Figure S1 b) shows the reflection and absorption for nanomagnetic arrays fabricated on an Au (250 nm) / SiO$_2$ (290 nm) substrate. Here, each nanomagnet is coated with a 4nm protective Al$_2$O$_3$ layer. Transmission into the substrate is inhibited by the gold under-layer, which enhances the absorption to 65\%.

Supporting Figure S2 shows the simulation absorption for nanomagnets in a square ASI geometry patterned on semi-infinite a) Si substrate and b) SiO$_2$ substrates. Nanomagnet dimensions of $L$ = 226 nm, $W$ = 78 nm and $P$ = 356 nm are used. The absorption at 633 nm is found to be 4$\%$ and 18$\%$ for Si and SiO$_2$ respectively. This is 16.25 and 3.61 $\times$ less absorption than the optimum Au (250 nm) / SiO$_2$ (290 nm) substrate in Supporting Figure S1 b).
\begin{figure*}[htbp]
    \centering
    \includegraphics[width=\textwidth]{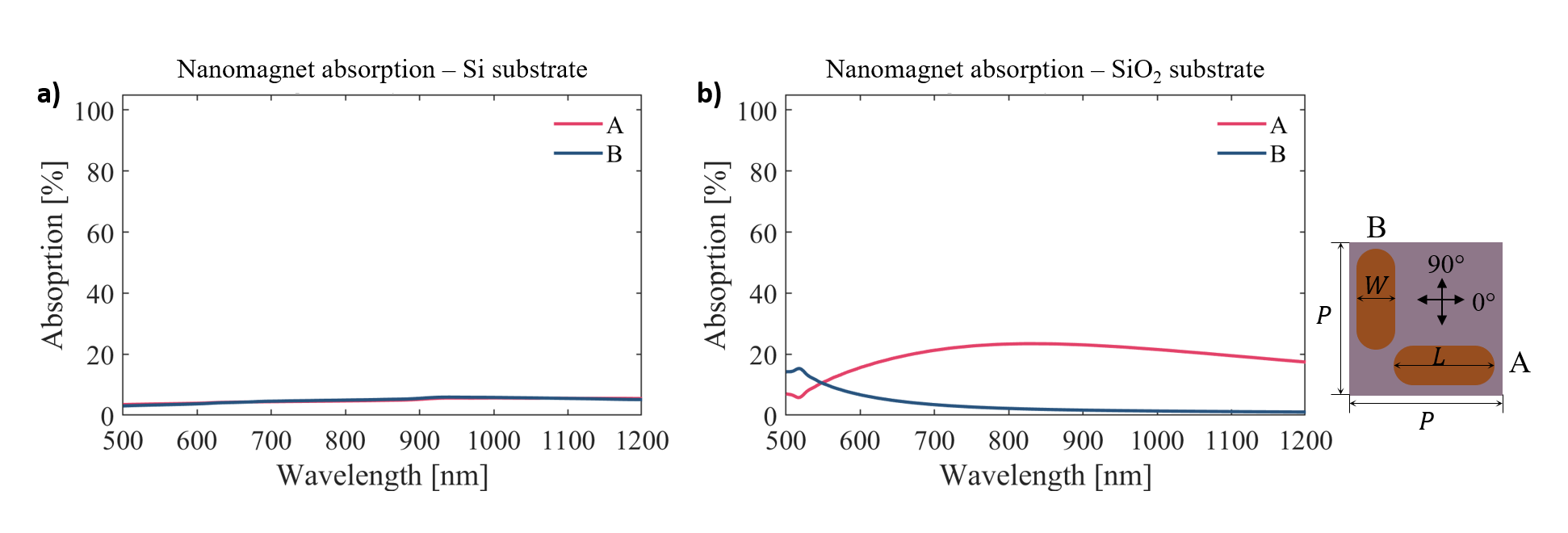}
    \caption*{\textbf{Supporting Figure S2.} Simulated absorption for nanomagnets patterned on semi-infinite a) Si substrate and b) SiO$_2$ substrate. Absorption at 633 nm is found to be 4$\%$ and 18$\%$ respectively. Nanomagnet dimensions of $L$ = 226 nm, $W$ = 78 nm, $t$ = 20 nm and $P$ = 356 nm are used.}
    \label{Si_absorption}
\end{figure*}

The linear optical response of the nano-structures was determined using the finite difference time domain (FDTD) technique (Lumerical FDTD). In these simulations, incident waves with different polarisations were applied to both individual structures and the periodic array. The simulations were simplified by including three layers of bar structures with rounded ends and a semi-infinite substrate. For single structure simulations, the absorption and extinction cross-sections were calculated by an analysis group of monitors located inside and outside of a total-field scattered-field source, respectively. A perfectly matched layer was applied in all directions to absorb incident light with minimal reflections. For array simulations, periodic boundary conditions were used in the plane of the substrate. In these simulations, the linear polarised waves of varying wavelength were incident from the air side. For the reflection calculation in Figure S1 a), we normalised to the reflection spectrum of a gold mirror. In the simulations, the complex refractive indexes of gold, silica, silicon, and aluminium oxide were taken from Palik models reported in ref\cite{edward1985handbook}. The refractive index of gold was taken from Johnson and Christy experimental data\cite{johnson1972optical}, and the refractive index of Py was taken from ref\cite{tikuivsis2017optical}.

Fourier transform infrared (FTIR) spectroscopy was used to characterise the linear optical response of the nanomagnetic particle arrays. The spectra were collected with a Bruker Hyperion 2000 FTIR microscope installed with a 15$\times$, NA = 0.4 metallic reflective objective. The reflection spectra were obtained by normalising the reflection curve from areas containing arrays and the neighbouring bare substrate against a reference spectrum taken from a gold mirror. Spectra were obtained in the ranges $500$–$1200$ nm using a silicon detector.

\subsection*{Supplementary note 2 - SEM images of nanomagnetic arrays used for FTIR measurements}
Supporting Figure S3 shows scanning electron micrographs (SEM) of nanomagnetic arrays patterned on top of a) Si / SiO$_2$ (300 nm) and b) Au (250 nm) / SiO$_2$ (290 nm) substrates which were used for FTIR measurements. Nanomagnetic array dimensions are consistent across both samples and are found to be a) $L$ = 226 nm, $W$ = 78 nm and $P$ = 356 nm and b)  $L$ = 223 nm, $W$ = 85 nm and $P$ = 353 nm. The patches observed in Supporting Figure S3 a) are attributed to residual PMMA. This does not significantly influence absorption in the nanomagnets as demonstrated by the switching fidelity observed throughout this work.
\begin{figure*}[htbp]
    \centering
    \includegraphics[width=\textwidth]{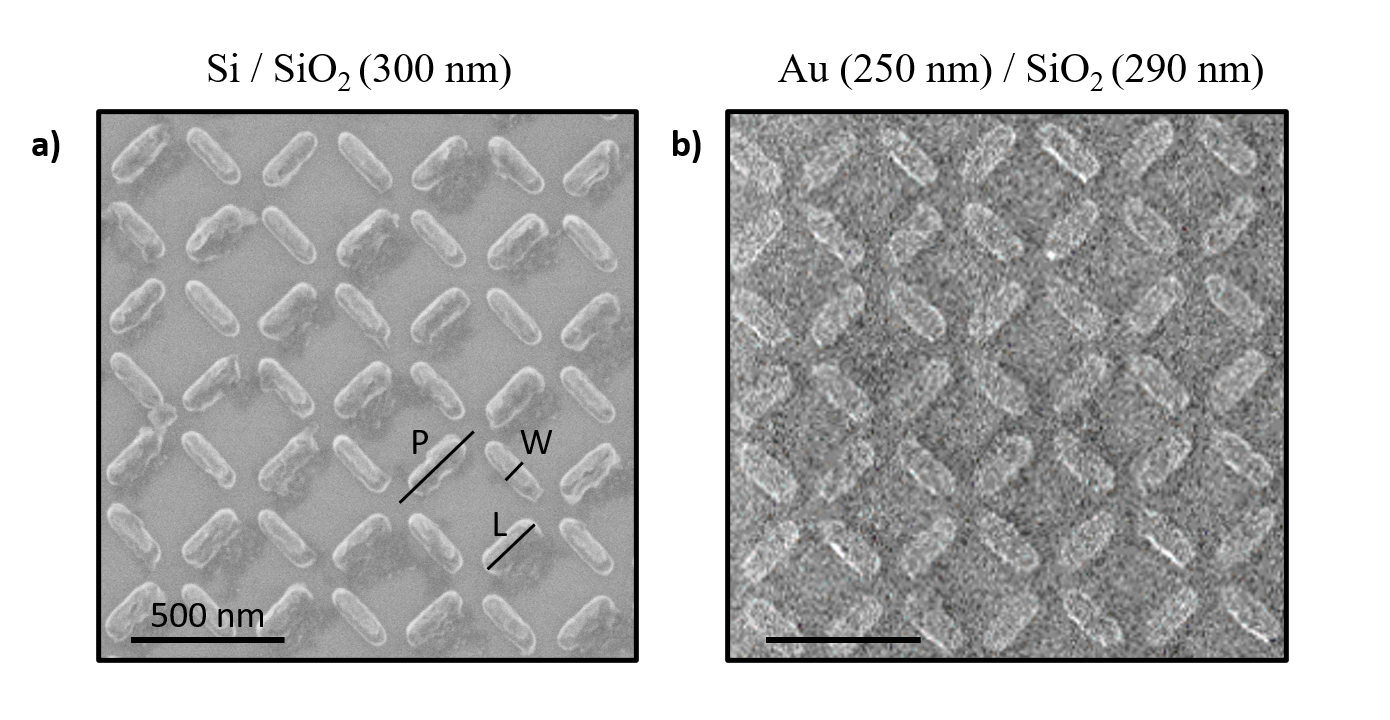}
    \caption*{\textbf{Supporting Figure S3.} SEM images of the nanomagnetic arrays patterned on top of a) Si / SiO$_2$ (300 nm) and b) Au (250 nm) / SiO$_2$ (290 nm) substrates. Dimensions are found to be a) $L$ = 226 nm, $W$ = 78 nm and $P$ = 356 nm and b)  $L$ = 223 nm, $W$ = 85 nm and $P$ = 353 nm.}
    \label{SEM}
\end{figure*}

\subsection*{Supplementary note 3 - Further information on the optical absorption mechanism}
Here we explore the localised surface plasmon (LSP) response of the permalloy (Py) nanomagnetic particles and argue that this is not the dominant absorption mechanism. Supporting Figure S4 shows the simulated absorption (panels a and c) and scattering (panels b and d) cross sections for a single nanomagnet on a silica substrate (panels a and b) and silicon substrate with a 300 nm silica layer (panels c and d). At the nanomagnet lengths and incident wavelengths explored in this work, low cross sections are observed. Furthermore, including the silicon substrate only has a weak influence on the LSP response. 

Supporting Figure S5 further explores nanomagnet optical absorption by varying the nanoarray unit-cell period (a) and silica thickness (c). We first fix the nanomagnet geometry and silica thickness, but vary array periodicity. Supporting Figure S5 a) shows the absorption for an ASI array with different periods overlayed with red and blue dashed lines indicating the grating diffraction orders described by $\lambda = n_{\mathrm{SiO2}}P$ and $\lambda = n_{\mathrm{SiO2}}P/\sqrt{2}$, where $P$ and $n_{\mathrm{SiO2}}$ are the period and the refractive index of the silica substrate, respectively. A diffraction order near a wavelength of 633 nm is observed for a periodicity = 500 nm, which decreases the absorption and should be avoided. Other periodicities and isolated nanomagnets perform well. Supporting Figure S5 b) shows the calculated absorption for a Py ASI array with a period $P = 600$ nm for a silica substrate (blue line) and silicon substrate with 300 nm silica layer (red line). The diffraction orders are blue-shifted for the silicon substrate with a silica layer which is attributed to interference within the silica layer. Figure S5 c) considers a fixed nanomagnet geometry and varying silica thickness, again exhibiting diffraction orders. Here, silica thickness is 300 nm, giving a resonance near the 633 nm wavelength of a He-Ne laser. It can be concluded that the nanomagnets have no clear plasmonic resonance that can provide strong absorption in the Py. Thus, for periodicity < 500 nm, the substrate interference phenomenon is the dominant absorption mechanism.

\begin{figure*}[htbp]
    \centering
    \includegraphics[width=\textwidth]{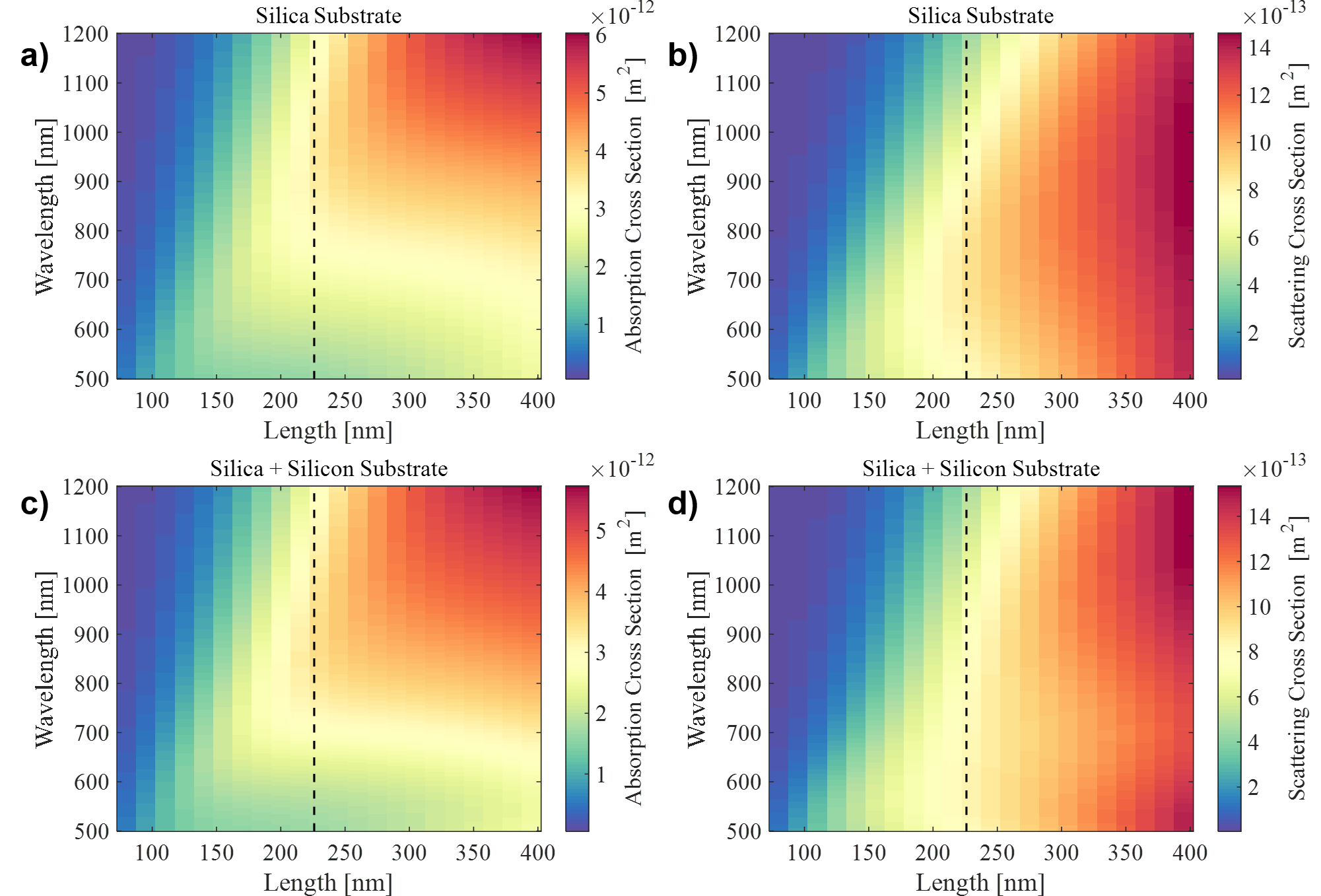}
    \caption*{\textbf{Supporting Figure S4.} Simulated LSP response of the nanomagnetic particles. (a and c) show the absorption cross sections and (b and d) show the scattering cross sections of a single nanomagnet particle. (a and b) are for a silica substrate and (c and d) are for silicon substrate with a silica layer (300 nm), respectively. The Silicon substrate has a very weak influence on the LSP response, which is quite weak in this wavelength range, with low cross sections.}
    \label{S1}
\end{figure*}

\begin{figure*}[htbp]
    \centering
    \includegraphics[width=\textwidth]{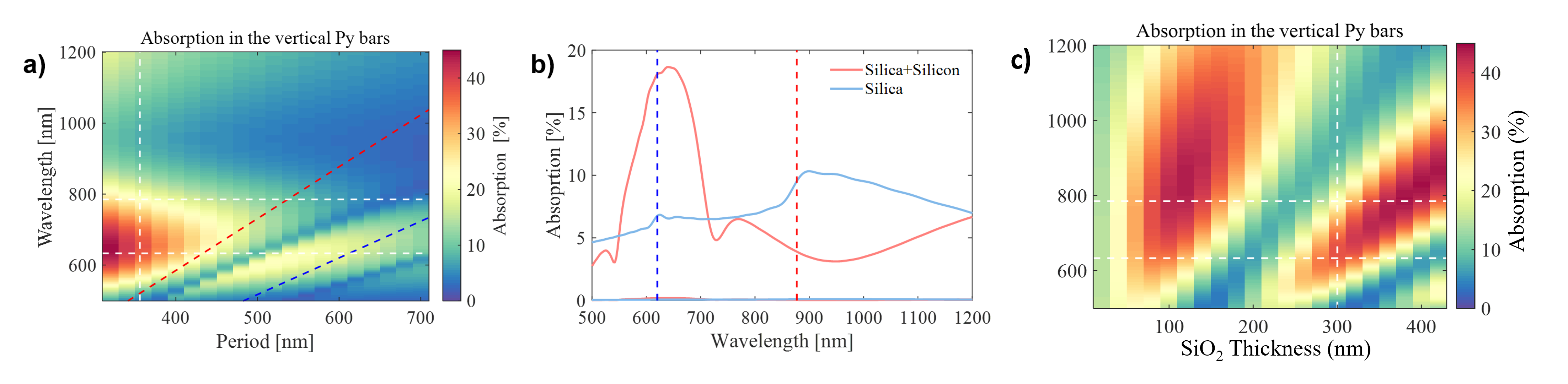}
    \caption*{\textbf{Supporting Figure S5.} The emergence of diffraction orders for large period values and their origin. a) Simulated optical absorption in the Py layers in the nanomagnet array for varying unit cell period. The vertical dashed line indicates the period of 356 nm used in experiments. The horizontal dashed lines indicate the wavelengths of 633 nm and 785 nm. The red and blue dashed lines indicate the grating diffraction orders, described by $\lambda = n_{\mathrm{SiO2}}P$ and $\lambda = n_{\mathrm{SiO2}}P/\sqrt{2}$, where $P$ and $n_{\mathrm{SiO2}}$ are the period and the refractive index of the silica substrate, respectively. b) Calculated absorption in Py for a nanomagnet array with $P$ = 600 nm for a silica substrate (blue line) and a silicon substrate with a silica layer (300 nm) (red line). The diffraction orders for the silica substrate match those shown in (a). The diffraction orders for the silicon substrate are clearly blue shifted, which we attribute to interference in the silica layer. c) Simulated optical absorption in the Py layer of nanomagnet array for varying SiO$_2$ thickness. Here equivalent dimensions to Figure S1 a) are used. }
    \label{S2}
\end{figure*}

\subsection*{Supplementary note 4 - Micromagnetic simulations of vertex energies}
\begin{figure*}[htbp]
    \centering
    \includegraphics[width=\textwidth]{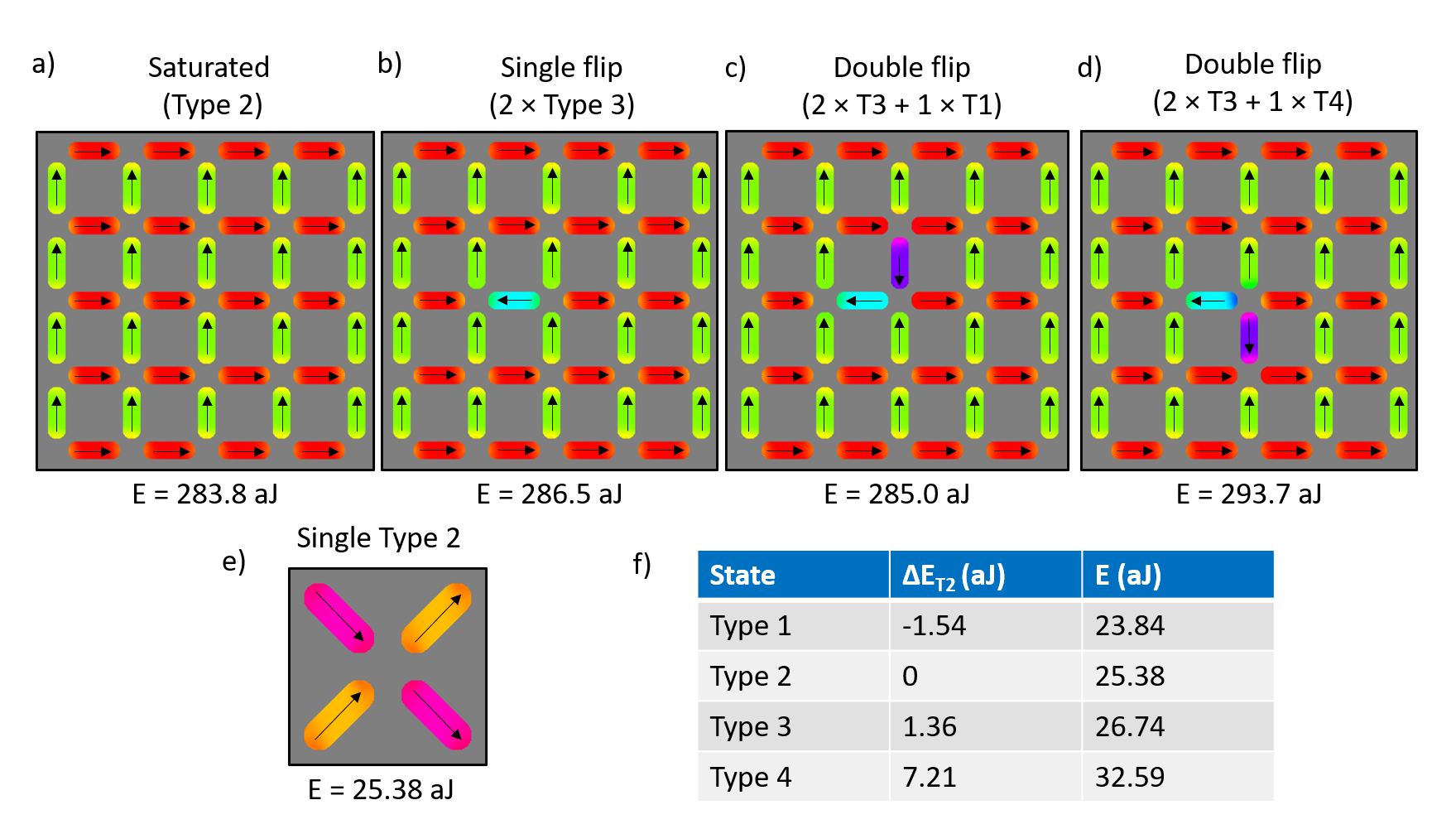}
    \caption*{\textbf{Supporting Figure S6.} Simulated microstate energies of a 5 $\times$ 5 vertex grid obtained from MuMax3. Nanomagnetic dimensions are $L$ = 226 nm, $W$ = 78 nm and $P$ = 356 nm, $t$ = 20 nm. Microstates and total energies of a) the saturated type 2 state. b) A single spin flip on a saturatd background. A double flip leaving a central c) type 1 vertex and d) type 4 vertex each with two type 3 vertices. e) Microstate and energy of a single type 2 vertex with periodic boundary conditions. f) Table of vertex energies expressed as relative to the T2 state ($\Delta$E\textsubscript{T2}) and absolute values (E). }
    \label{mumax}
\end{figure*}

Micromagnetic simulations were used to obtain an estimate of each vertex energies for the nanomagnetic arrays explored in this work. Supporting Figure S6 shows four microstates in a) a saturated type 2 state, b) a single switch on a saturated background leaving 2 type 3 monopoles. c,d) A further spin flip giving c) 2 x type 3 and 1 x type 1 and d) 2 x type 3 and 1 x type 4. From this the relative energies of each state are calculated. e) shows the microstate and energy of a single type 2 vertex with periodic boundary conditions. f) Shows the relative (compared to the type 2 state) and absolute energies for each vertex type. The temperature is set to 0 K making the energies a likely overestimate. Nanomagnetic dimensions are $L$ = 226 nm, $W$ = 78 nm and $P$ = 356 nm, $t$ = 20 nm.

\subsection*{Supplementary note 5 - Scanning beam switching on Au/SiO$_2$ substrates}
\begin{figure*}[htbp]
    \centering
    \includegraphics[width=0.5\textwidth]{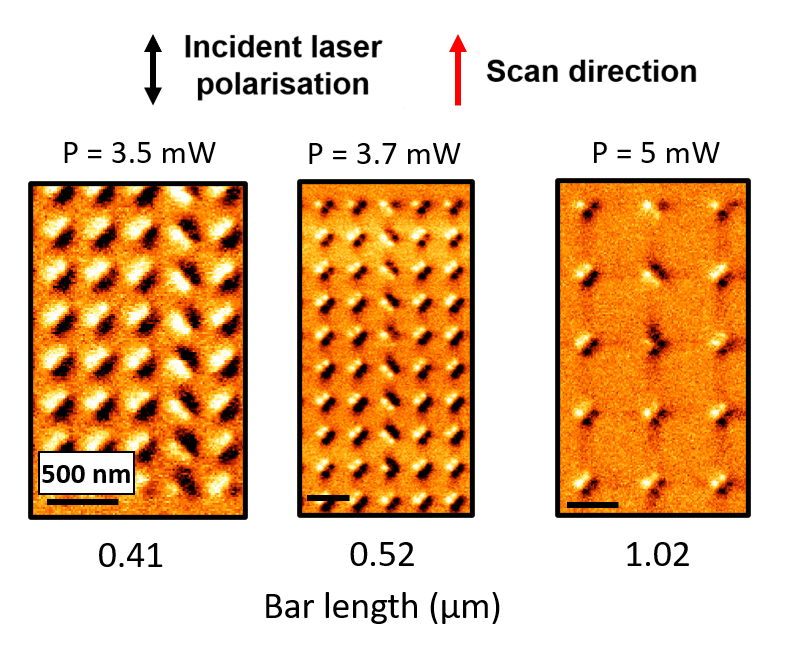}
    \caption*{\textbf{Supporting Figure S7.} MFM images of switching of nanomagnets patterned on an Au/SiO$_2$ substrate. Bar lengths up to 1.02 \textmu m are reversed. Here the scan direction and polarisation are parallel. A higher power is required for longer nanomagnets}
    \label{Au_scan}
\end{figure*}

Supporting Figure S7 shows MFM images of a range of ASI arrays with increasing nanomagnet lengths patterned onto a Au / SiO$_2$ substrate. Here, the scan direction and polarisation are parallel. All nanomagnet lengths up to 1.02 \textmu m can be reversed, demonstrating the efficacy of the technique when combined with a substrate which inhibits transmission.

\subsection*{Supplementary note 6 - Magnetometry of nanomagnetic arrays}
\begin{figure*}[htbp]
    \centering
    \includegraphics[width=\textwidth]{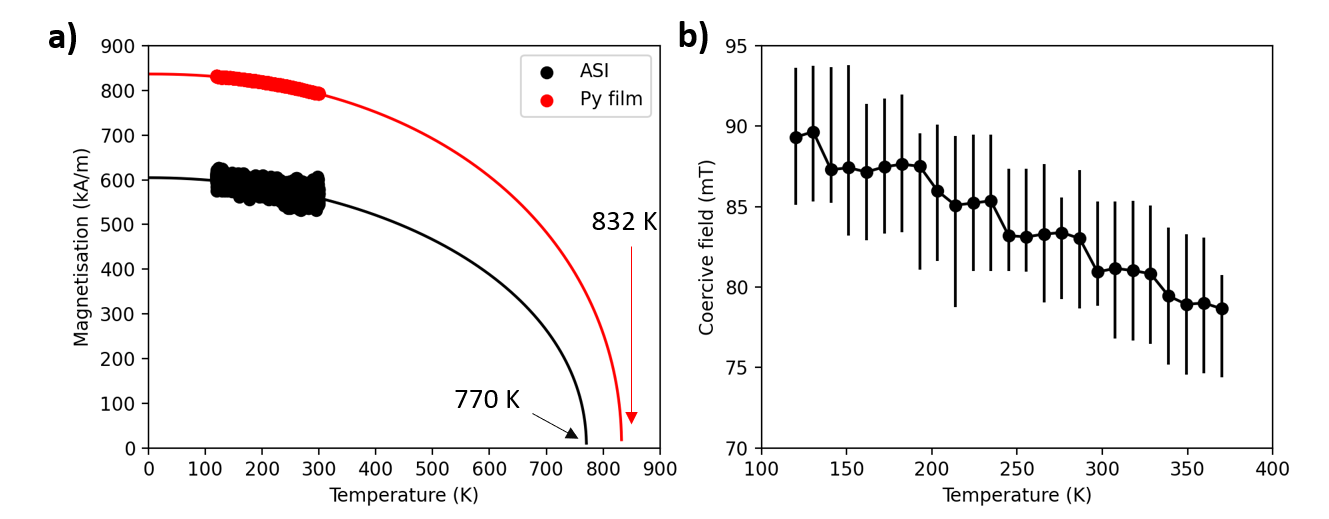}
    \caption*{\textbf{Supporting Figure S8.} Showing a) Magnetisation as a function of temperature between 120 K and 380 K for a 2 mm x 2 mm ASI array and a 6.5 mm x 5.5 mm thin film deposited at the same time. Both datasets are fitted to M(T) = M(0)(1-(T/T\textsubscript{c})\textsuperscript{$\alpha$})\textsuperscript{$\beta$} where T\textsubscript{c} is the Curie temperature, M(0) and M(T) are the magnetisation at 0 K and T K respectively and $\alpha$ and $\beta$ are empirical constants where $\alpha$\textsubscript{ASI} = 2,  $\alpha$\textsubscript{film} = 2.18 , $\beta$\textsubscript{ASI} = 0.476,  $\beta$\textsubscript{film} = 0.476. T\textsubscript{c} is estimated to be 770 K and 832 K for ASI and the reference film respectively. Fits to conventional bloch law yield unrealistic high T$_c$ b) Coercive field field of the ASI array as a function of temperature. Error bars indicate the start and end of switching. The coercive field (H\textsubscript{c}) reduces by 10.7 mT from 120 - 370 K. Magnetic field is applied at 45$^{\circ}$ to both subsets of bars.}
    \label{mvst}
\end{figure*}

Supplementary Figure S8 a) shows the temperature dependence of magnetisation of a 2 mm x 2 mm square ASI array (Py) and Py thin film. The sample is the same one used for the FTIR measurements in Supporting Figure S1 a). A $\sim$10$\%$ reduction in magnetisation occurs between 120 - 380 K. Also shown in extrapolated fits to the equation M(T) = M(0)(1-(T/T$_c$)$^\alpha$)$^\beta$) \cite{kuz2005shape}. T$_c$ is found to be 770 K and 832 K for ASI and the thin film respectively. Fits to conventional bloch law M(T) = M(0)(1-T/T$_c$)$^{3/2}$ yield unrealistically high values of T$_c$. Nevertheless, it is unlikely that the temperatures reached during CW exposure reach T$_c$. Supplementary Figure S8 b) shows the coercive field of the nanomagnets as a function of temperature from 120 - 370 K. Error bars indicate the start and end of switching arising from the distribution of coercive fields across the sample. Magnetic field is applied at 45$^{\circ}$ to both subsets of bars. The coercive field reduces by 10.7 mT in this range indicating a high T$_c$ of the samples.

\subsection*{Supplementary note 7 - Ni$_{50}$Fe$_{50}$ and Co nanomagnetic array exposure} 
Supporting Figure S9 shows writing of a,b) Ni$_{50}$Fe$_{50}$ (20 nm thickness) and c) attempted writing of a Co (8 nm thickness) nanostructures patterned on an Au/SiO$_2$ substrate. Co dimensions are L = 320 nm, W = 110 nm, P = 385 nm.  Ni$_{50}$Fe$_{50}$ possesses the necessary band structure for OISTR to take effect whereas Co does not. Ni$_{50}$Fe$_{50}$ nanowires with length $\leq$ 490 nm are switched at 3 mW power. This is a lower power and broader dimension set than observed in Py. Conversely, no switches are observed in Co for powers up to 8 mW displayed and 30 mW (not shown). This is consistent across a broad range of nanomagnet dimensions. These results indicate that the multi-species nature of NiFe alloys which allow for the necessary band structure plays a role in the observed switching. Lower contrast in Supporting Figure S8 b is due to the necessity of using a low-moment MFM tip for this thickness of Co.

\begin{figure*}[htbp]
    \centering
    \includegraphics[width=\textwidth]{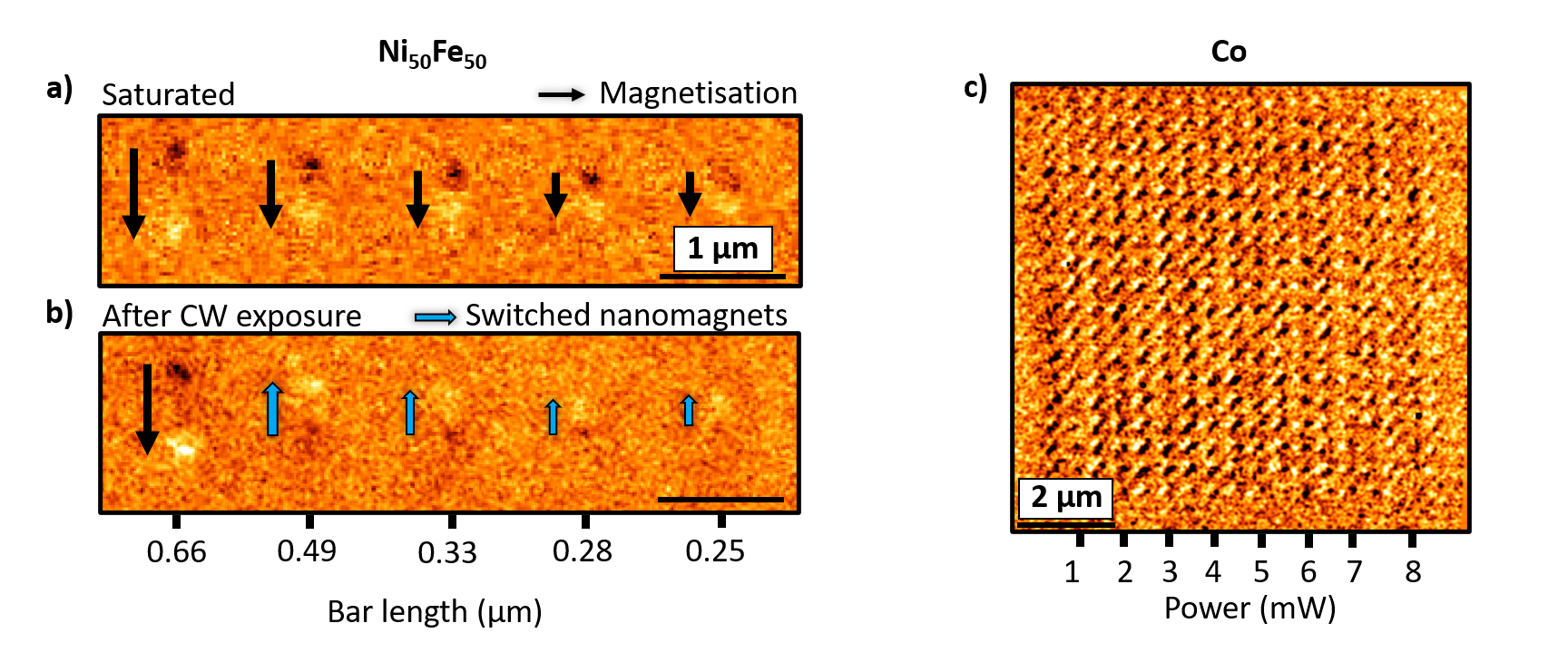}
    \caption*{\textbf{Supporting Figure S9.} mFM images of Ni$_{50}$Fe$_{50}$ in a) a saturated state and b) after exposure to a $\lambda$ = 633 nm laser with 3 mW power. Polarisation is parallel to the bar long axis. Bars with length $\leq$ 0.49 \textmu m switch. c) Shows MFM image of a Co array with nanomagnet dimensions of L = 320 nm, W = 110 nm, P = 385 nm after exposure to a $\lambda$ = 633 nm laser with powers ranging from 1-8 mW. No switches are observed. Here we use a low moment MFM tip for imaging to avoid MFM-tip writing\cite{gartside2018realization}.}
    \label{COFENI}
\end{figure*}

\end{document}